\documentclass[twocolumn]{aastex7}

\usepackage{amsmath}

\newcommand{\nt}{\textcolor{black}}
\newcommand{\ntt}{\textcolor{black}}

\begin{document}

\title{Silicate Sundogs: Probing the Effects of Grain Directionality in Exoplanet Observations}

\author[0000-0003-0814-7923]{Elijah Mullens}
\email{eem85@cornell.edu}
\affiliation{Department of Astronomy and Carl Sagan Institute, Cornell University, 122 Sciences Drive, Ithaca, NY 14853, USA}

\author[0000-0002-8507-1304]{Nikole K. Lewis}
\email{nkl35@cornell.edu}
\affiliation{Department of Astronomy and Carl Sagan Institute, Cornell University, 122 Sciences Drive, Ithaca, NY 14853, USA}

\begin{abstract}
Crystalline ice in Earth's atmosphere can produce spectacular phenomena due to orientation-dependent attenuation, such as sun dogs and halos, providing diagnostics of the external processes acting on the aerosol grains. Crystalline mineral aerosols, such as quartz (SiO$_2$) and enstatite/forsterite (MgSiO$_3$/Mg$_2$SiO$_4$), have long been predicted to form in hot Jupiter atmospheres with JWST MIRI LRS verifying \nt{the existence of crystalline quartz} observationally. Due to the strong horizontal winds \nt{($\sim$ 1 -- 5 km s$^{-1}$)} and small aerosol grains ($<1$~\textmu m) found in hot Jupiter atmospheres, we show that aerosols could be mechanically aligned with the winds. We then derive directional-dependent optical properties of quartz, enstatite, and forsterite and model transmission and emission spectra assuming random and mechanically aligned orientations, finding that the orientation of \nt{all three} crystalline aerosols can impart $\geq$ 100 ppm differences in observed spectra \nt{(8 -- 12 \textmu m)}. We run retrievals on JWST MIRI LRS transmission and emission data of WASP-17b and find that directionality alone cannot physically explain the transmission data, pointing towards polymorphs or insufficient lab data, and find \nt{weak} hints of directionality \nt{(1.0 -- 1.3$\sigma$)} in the emission data. This work demonstrates the power of JWST MIRI LRS in detecting aerosol directionality with future observations, and a technique by which to probe how aerosols interact with atmospheric dynamical processes. \nt{To foster the exploration of aerosols in exoplanet data, the open-source code \texttt{POSEIDON} has been updated (v1.3.1) to include 144 new directional and temperature aerosols with precomputed optical properties, alongside new aerosol models.}
\end{abstract}


\section{Introduction} \label{sec:intro}

The alignment of crystalline aerosols in planetary atmospheres can produce a myriad of spectacular optical phenomena due to their interaction with light. As hexagonal ice crystals fall throughout Earth's atmosphere they align horizontally, generating subsuns and light pillars through reflection and sundogs and `fire rainbows' (circumhorizontal arcs) through refraction. Parhelic circles form in the rare case of reflective vertically aligned crystals, while the 22$^{\circ}$ halo is due to randomly oriented refractive crystals. Crown flashes form when ice crystals orient themselves with the electric field above thunderclouds. These optical phenomena are not only beautiful, but are also a powerful diagnostic of the local atmospheric processes where the crystals form.  

Similarly, the polarization of starlight by the Interstellar Medium (ISM) and protoplanetary disks has been attributed to the alignment of non-spherical dust grains by a variety of physical processes \citep[e.g. see reviews][]{Lazarian2007, Andersson2015, Hoang2023}. ISM and protoplanetary dust grains form via the same condensation pathways that are thought to occur in substellar atmospheres \citep{Sargent2009,Visscher2010}. Silicate-based condensate clouds have been detected in the atmospheres of brown dwarfs \citep[e.g.][]{Cushing2006, Burningham2021}, planetary mass companions \citep[][]{Miles2023}, and hot-Jupiter exoplanets \citep[e.g][]{Grant2023, Dyrek2024, Schlawin2024,Ingles2024}. These silicate-based aerosols, or dust, in exoplanetary atmospheres will be subjected to many of the same external forces as dust grains in ISM and protoplanetary environments such as magnetic fields, radiation, and winds. 

\citet{Gold1952} first proposed that the alignment of dust grains in the ISM could be due to the dynamical interaction of dust grains and the gas flows around them. This so-called ``mechanical alignment" of dust grains in the ISM leads to an orientation of the dust grain long axis with the relative velocity between the gas and dust. However, for the low density gas environment of the ISM ($\sim$1~atom~cm$^{-3}$), studies have shown that alignment of grains due to interaction with magnetic fields \citep[e.g.][]{Davis1951,Lazarian2007} and radiation pressure \citep[e.g.][]{Harwit1970,Lazarian1995} are likely the dominate grain alignment processes. In the higher density gas environments of protostellar and protoplanetary disks ($\sim 10^{-8}$~g~cm$^{-3}$), mechanical alignment of dust grains contributes more significantly with several studies noting polarization signals indicative of alignment with respect to the background gas flow \citep[e.g.][]{Kataoka2019, Hoang2022}. In the high density gas environments of Gas Giant planet atmospheres ($\sim10^{-3}$~g~cm$^{-3}$), one would expect significant coupling between atmospheric gas and ``dust" (aerosols) that would give rise to preferential alignment of non-spherical grains. Hot Jupiter exoplanets provide not only a dense gas environment rich with aerosols, but also strong zonal jets with nearly supersonic winds \citep[see review][]{Showman2020}, a regime ripe for the mechanical alignment of ``dust" grains.   

Recent work by \citet{Hoang2023} has considered the alignment of ``dust" grains in hot Jupiter atmospheres via radiative torques and possible magnetic fields. 
\nt{In their study, \citet{Hoang2023} primarily focus on these effects in the low-density upper atmospheres of hot Jupiters ($n_H<10^{13}$~cm$^{-3}$, or $p<10$~\textmu bar) where the anisotropic radiation field from the host star can drive radiative torques (RATs) and potential alignment of silicate grains with the magnetic field (B-RAT). In deriving the minimum grain size for alignment by RATs/B-RAT in a hot Jupiter atmosphere, \citet{Hoang2023} show that only large grains ($a>>1$~\textmu m) are susceptible to such alignment at altitudes/pressures relevant to exoplanet transmission and emission observations at visible and infrared wavelengths (1~bar to 0.1~mbar). Small grains ($a<1$~\textmu m) may still be susceptible to alignment with any ambient magnetic field if they are composed of a paramagnetic material (e.g. amorphous silicates or have iron inclusions), but the maximum grain size for external alignment with the magnetic field scales with the gas density and would require $a<10^{-4}$~\textmu m at hot Jupiter photospheric densities ($n_H\sim10^{16}$~cm$^{-3}$) \citep{Hoang2022b}.}

\nt{JWST MIRI observations of Hot Jupiters \citep[e.g][]{Grant2023, Ingles2024} have detected $\sim10^{-2}$~\textmu m crystalline quartz (SiO$_2$) aerosols in $P<1$~mbar atmospheric regions. Following \citet{Hoang2022b, Hoang2023} these aerosols are unlikely to be aligned by radiative torques or magnetic fields, but these studies did not consider possible multi-dimensional effects from strong day-night temperature contrasts and the associated horizontal winds that are generated \citep[e.g.][]{Showman2002}. 
}Here we consider the potential for ``mechanical alignment" of crystalline silicate-based aerosols in hot Jupiter due to coupling with the $\sim$1--5~km~s$^{-1}$ horizontal winds present in these atmospheres. 
In the sections that follow we first explore the potential for mechanical alignment of small grains in exoplanet atmospheres, then estimate the observability of silicate-based aerosol grain alignment at mid-infrared wavelengths accessible with JWST MIRI transmission and emission spectroscopy (5-14 \textmu m).

\section{Mechanical Alignment of Small Grains in Exoplanet Atmospheres} \label{sec:mechanical-alignment}
Here we explore the possible mechanical alignment of aerosols in exoplanet atmospheres following a similar explorations for proptoplanetary disks and the ISM \citep[e.g.][]{Kataoka2019, Birnstiel2010, Lazarian1994}. \nt{\citet{Tsuchiyama1998} found that minerals that form in hot Jupiter-like environs form as spheroids, elongated alongside one axis. Therefore, we assume} that aerosols are prolate spheroids and focus on gas and aerosol velocities in the zonal (longitudinal) direction \nt{(see \S \ref{sec:aerosol-shape-appendix} for more details on aerosol shape)}. 
\nt{In nominal hot Jupiter atmospheres (1000~K$<T_{eq}<$2000~K) average zonal wind speeds are expected to be on the order of 1--5 km~s$^{-1}$ while average vertical wind speeds in the stably stratified photospheric regions (1~bar to 0.1~mbar) will be on the order of 10~m~s$^{-1}$ \citep[e.g.][]{Showman2002, Showman2020, Parmentier2021, Roth2024}.}
The coupling between aerosols and gas in a planetary atmosphere (or protoplanetary disk, ISM, etc.) can be described by the dimensionless {\it Stokes number}:
\begin{equation}
    St=\frac{\tau_{stop}}{\tau_{eddy}}
\end{equation}
Where $\tau_{stop}$ is the particle stopping time and $\tau_{eddy}$ is the eddy overturn timescale. 

The value of $\tau_{stop}$ will depend on the {\it Knudsen number} ($Kn$) and {\it Reynolds number} ($Re$) which are tied to the average aerosol particle size and the gas viscosity. Here we assume that the aerosol grains are submicron sized silicate grains in line with recent determinations from JWST observations of hot Jupiters \citep[e.g.][]{Grant2023,Ingles2024, Dyrek2024}. The Knudsen number is defined as the ratio between the mean-free path of the gas molecules ($\lambda_{mfp}$) and the aerosol particle size \citep[e.g.][]{Rossow1978, Birnstiel2010}. For typical hot Jupiter atmospheric conditions near the 1~mbar pressure level, we would expect $\lambda_{mfp}$ to be on the order of hundreds to thousands of \textmu m, which is significantly larger than our assumed submicron sized aerosol grains ($Kn>>1$). 

Although a variety of formulations for estimating the \ntt{aerosol particle} Reynolds number exist \citep[e.g.][]{Rossow1978, Birnstiel2010}, at its core \ntt{$Re_{ap}$} is a ratio between the aerosol advection and the viscosity of the gas. 
\nt{We estimate the \ntt{aerosol particle Reynolds-number $Re_{ap}=2\rho a v/\eta$}, where $\rho$ is the atmospheric density in the region of interest (1-100~mbar, T$\sim$1500~K, $\rho\sim10^{-4}$kg~m$^{-3}$), $a$ is the particle radius, $v$ is the particle velocity, and $\eta$ is the atmospheric dynamic viscosity in the region of interest. For a hydrogen-dominated hot Jupiter atmosphere $\eta$ can be estimated to be on the order of 2.5$\times10^{-5}$~Pa~s according to formula presented in \citet{Parmentier2013} from \citet{Rosner2000}. 
Assuming submicron sized aerosol grains moving at speeds similar to the background flow we estimate \ntt{$Re_{ap}<<1$}.}
This places the hot Jupiter aerosols considered in this study in the {\it Epstein} flow regime \citep{Epstein1924, Birnstiel2010}. Most previous studies of aerosol transport in Hot Jupiters focused on the Stokes flow regime \citep[e.g][]{Parmentier2013}, but the Epstein flow regime is commonly considered in the protoplanetary disk community \citep[e.g.][]{Birnstiel2010}. However, the aerosol stopping time ($\tau_s$) estimated for both the Epstein and Stokes flow regimes are both similarly small ($\tau_s<<1~s$). 

In estimating the eddy turn-over time {$\tau_{ed}$}, previous studies have focused primarily on vertical transport and settling \citep[e.g. $K_{zz}$][]{Ackerman2001}. 
\nt{In the stably stratified photospheric regions of Hot Jupiter atmospheres we would expect that bulk vertical mixing would occur on timescales of order an Earth day or longer, with localized updrafts and downdrafts where vertical mixing timescales could be on the order of an hour \citep[e.g.][]{Parmentier2013,Komacek2019}.}
Similarly, if we consider the timescale for transport in the day-to-night (zonal) direction ($K_{xx}$) in a hot Jupiter atmosphere with winds approaching (or exceeding) the speed of sound, we find $\tau_{ed}$ on the order of a an Earth day. In any case, we can say with confidence that $\tau_{ed}>>\tau_{s}$ ($St<<1$) and that the submicron-sized aerosols should be well-coupled (follow streamlines) to the atmospheric gas flow. Given these conditions, aerosols in hot Jupiter atmospheres are very likely to experience Gold-type grain alignment \citep[e.g.][]{Lazarian1994}.

\begin{figure*}
\center
    \includegraphics[width=0.99\textwidth]{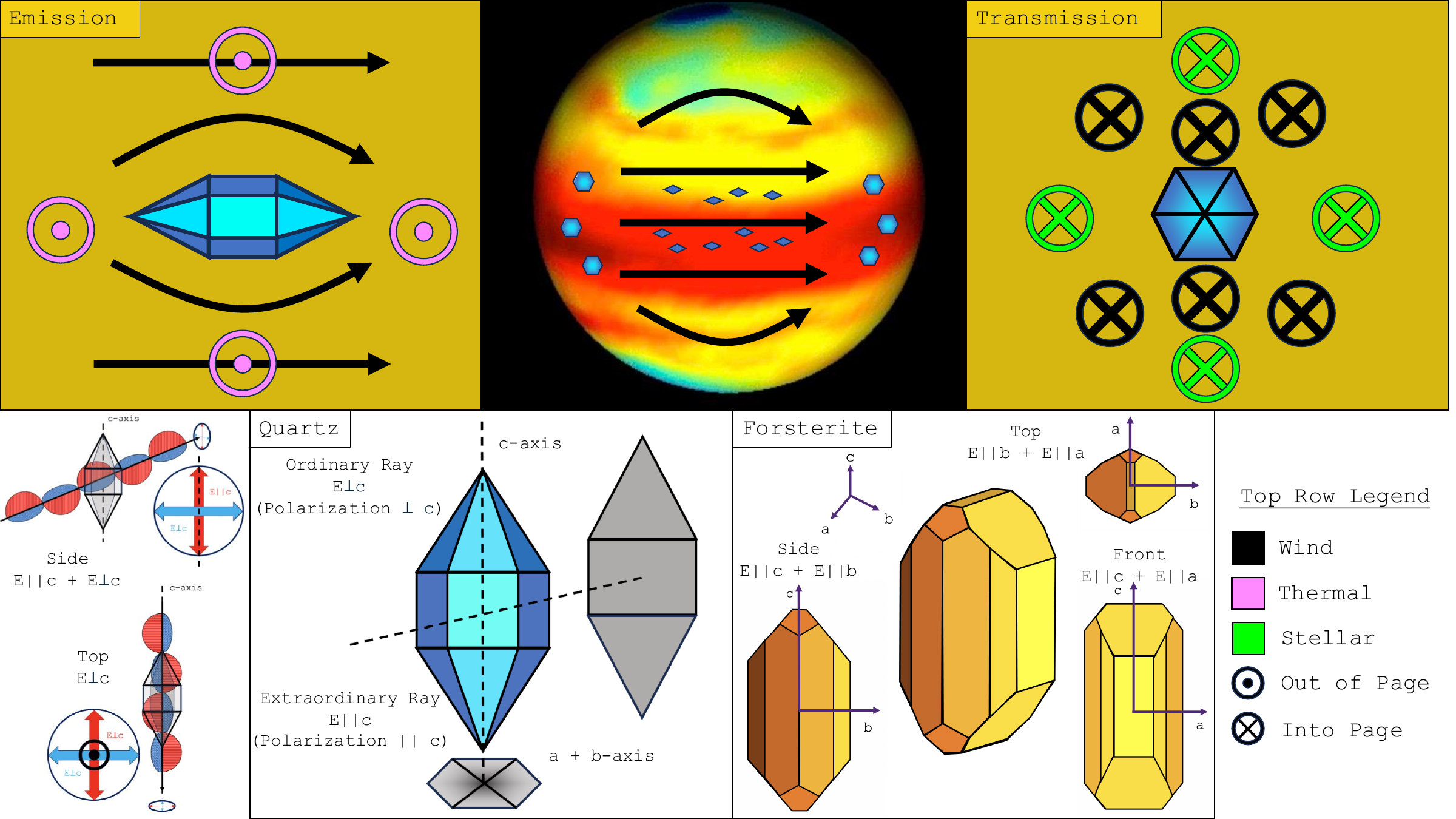}
    \caption{Crystalline silicate grains predicted to form in hot Jupiters grow in a non-spherical shape and interact with mechanical forces (i.e., wind) that will preferentially orient the crystals. \textbf{Top}: Diagram showing how a quartz crystal would be oriented relative to an observer (assuming mechanical alignment with winds) for emission and transmission geometries. In emission, observed outgoing thermal radiation will interact with the `sides' of the crystal while in transmission incident starlight will interact with the `top' of the crystal. \textbf{Bottom}: Quartz is a uniaxial crystal with a single optical axis (c), meaning its radiative properties are determined by the extraordinary (polarization $\parallel$ c, or E $\parallel$ c) and ordinary rays (polarization $\perp$ c, or E $\perp$ c), each with their own set of refractive indices. When light propagates through the side of the crystal, the radiative properties will be determined by both ordinary and extraordinary rays, and when light propagates top to bottom the radiative properties will be determined only by the ordinary ray. Assuming a random orientation of crystals leads to a weighting of 2/3 ordinary + 1/3 extraordinary.  Forsterite is a biaxial crystal with two optical axes and three sets of refractive indices (polarization $\parallel$ c, b, a). Assuming light propagates along the a-axis (`side'), the radiative attenuation will be determined by the polarization $\parallel$ c and b refractive indices (and similarly, each axis is a combination of two sets of refractive indices). Assuming a random orientation of crystals leads to a 1/3 + 1/3 + 1/3 weighting. Quartz diagram adapted from the Quartz page (\url{http://www.quartzpage.de/gen_struct.html}). Forsterite diagram adapted from \citet{Beno2020}.}    
    \label{fig:crystal-art}
\end{figure*}

\section{Aerosol Opacity Effects}

Here we give a primer on how radiative anisotropy arises in crystalline aerosols, and how anisotropy propagates to derived cross sections.

\subsection{Crystalline Aerosol Anisotropy}\label{sec:crystalline-aerosol-anisotropy}


Aerosols are largely classified as either amorphous, which have no crystal structure and didn’t crystallize due to cooling too fast (‘quenching’) \nt{or because they are organic in nature (e.g., organic hazes)}, or crystalline, which crystallize and have regular structure. It is expected that silicates condense in a crystalline state at high temperatures \citep{Tsuchiyama1998} and amorphous states at lower temperatures \citep{Fabian2000}, where amorphous aerosols can transition to a crystalline state by being exposed to high temperatures for a set period of time (i.e. annealing on the timescale of hours) \citep{Gail1998, Fabian2000}. Amorphous and crystalline aerosols interact with radiation differently. Amorphous aerosols always interact with light isotropically, and due to the disordered arrangement of molecules found in amorphous materials, vibrational modes become `smeared' and impart broad absorption features with no fine structure \citep{HenningMutschke1997}. Additionally, amorphous absorption bands are much less sensitive to temperature variations. 

Given the high temperatures predicted on the daysides and terminators of hot Jupiters ($\sim$ 1000-2000 K), we expect the dominant form of silicate aerosols to be crystalline, either through direct condensation or subsequent processing (e.g., annealing). Indeed, retrieval analysis of WASP-17b's JWST MIRI LRS transmission spectra found tentative evidence that crystalline SiO$_2$ was preferred over amorphous \citep{Grant2023}.

Crystalline aerosols have absorption features with fine structure (`sharp' absorption features) that are more sensitive to temperature variations. Additionally, they can interact with light isotropically or anisotropically, where how light interacts with a crystalline aerosol's crystallographic axes (labeled a, b, and c in Figure \ref{fig:crystal-art}) depend on its intrinsic crystal system and resultant crystal shape through its growth habit (see Appendix \S \ref{sec:crystalstructure} for more details).

Silicates and their polymorphs display a diverse range of crystal systems. $\alpha$ quartz (SiO$_2$) is a trigonal/\nt{rhombohedral} crystal system, which has a single optical axis (i.e., uniaxial). Uniaxial crystals are anisotropic and have two sets of refractive indices: polarizations of light parallel to the c-axis (extraordinary, A-mode, E$\parallel$c) and polarizations perpendicular to the c-axis (ordinary, E-mode, E$\perp$c). Forsterite (Mg$_2$SiO$_4$) and enstatite (MgSiO$_3$) both have orthorhombic crystal systems with two optical axes (i.e., biaxial). Biaxial crystals are anisotropic and have three defined sets of refractive indices that correspond to light being polarized relative to three optical directions (which, for orthorhombic crystals, align with the crystallographic axes: E$\parallel$c, E$\parallel$b, E$\parallel$a). $\beta$ cristobalite (SiO$_2$), a high temperature polymorph of SiO$_2$, is a cubic crystal system and therefore optically isotropic.  

To derive the optical properties of aligned crystalline aerosols, we have to consider how the crystal is oriented relative to the direction of light propagation, and which refractive indices (which are defined relative to polarization, not propagation) to use. For uniaxial crystals, light propagating parallel to the c-axis (hereafter, called `Top') will always have polarization perpendicular to the c-axis and therefore have attenuation determined solely by the ordinary refractive indices. Light propagating perpendicular to the c-axis (hereafter called `Side') can have polarizations perpendicular and parallel to the c-axis and therefore have attenuation determined by both ordinary and extraordinary refractive indices (see Figure \ref{fig:crystal-art}, bottom left). Because of this, when weighing the optical properties of a system of randomly oriented crystals the extraordinary ray receives a 1/3 weight while the ordinary ray receives a 2/3 weight \citep[e.g.,][]{Zeidler2013}. 

For biaxial crystals, light propogating parallel to each axis will have attenuation determined by two sets of refractive indices; for example, light propagating along the b-axis with a polarization parallel to the c-axis will be attenuated via the E$\parallel$c refractive indices while light with polarization parallel to the a-axis will be attenuated via the E$\parallel$a refractive indices. For the remainder of the paper, we will refer to light propagating along the b-axis as `Front', a-axis as `Side', and c-axis as `Top' for biaxial crystals. For randomly oriented biaxial aerosols, each set of refractive indices receives a 1/3 weight. 

\nt{Crystalline uniaxial and biaxial crystals that form or grow in non-spherical shapes (see \S \ref{sec:aerosol-shape-appendix}), and therefore can be expected to be mechanically aligned to the wind (see \S \ref{sec:mechanical-alignment}) provide a mechanism by which to probe aerosol and gas coupling due to orientation-dependent attenuation. In the next section, we derive orientation-dependent cross sections.}

\subsection{Refractive Indices and Cross Sections}\label{sec:cross_sections}


\begin{figure*}
    \includegraphics[width=0.99\textwidth]{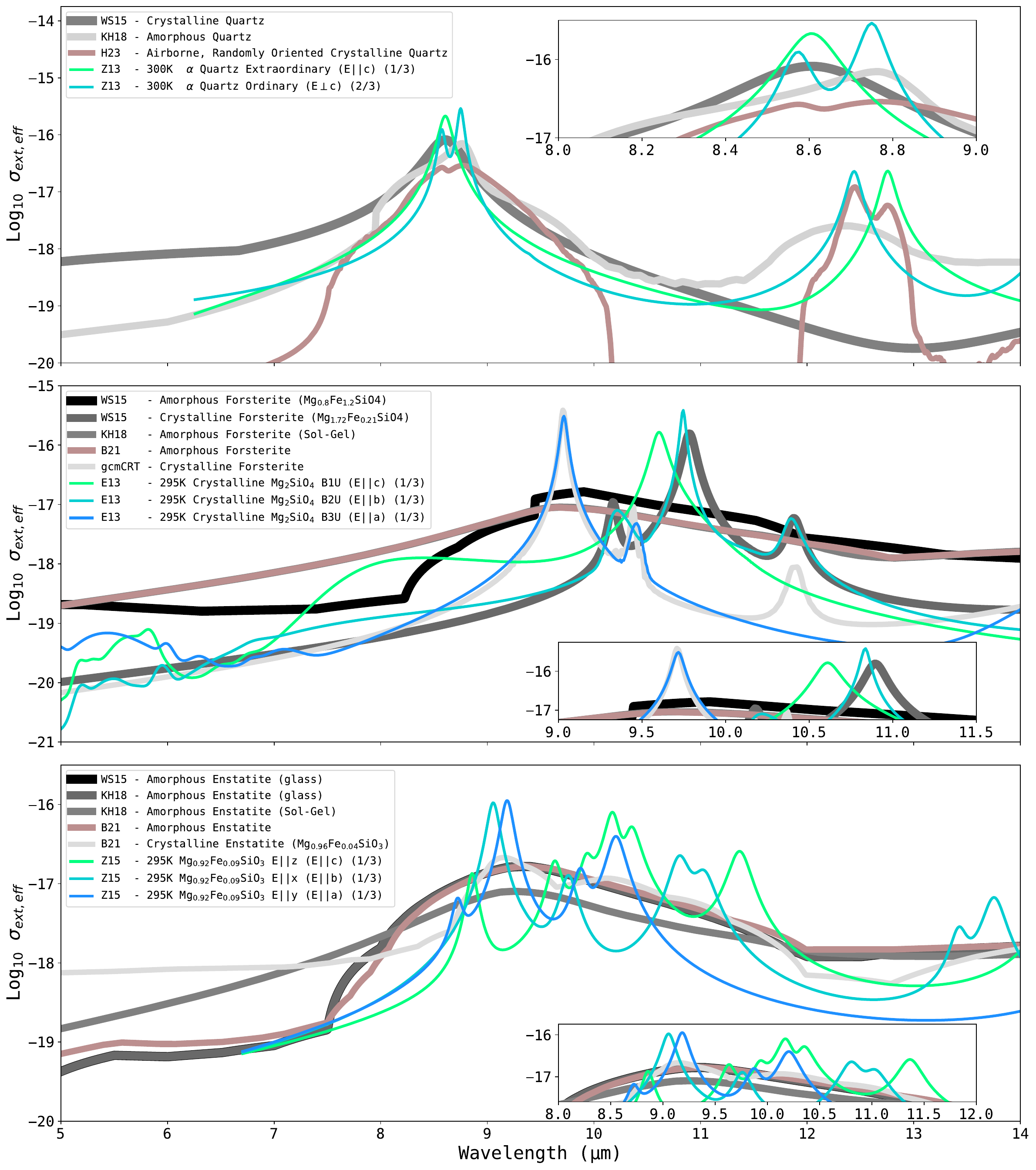}
    \caption{Comparison of effective extinction cross sections ($\sigma_{\mathrm{ext,eff}}$, mean particle radii = 0.01 \textmu m with a lognormal distribution [log width of 0.5]) computed from refractive indices used extensively in exoplanet literature vs those computed from room-temperature directional-specific refractive indices. \textbf{Top}: Comparison of quartz (SiO$_2$) cross sections. \textbf{Middle}: Comparison of forsterite (Mg$_2$SiO$_4$) cross sections. \textbf{Bottom}: Comparison of enstatite (MgSiO$_3$) cross sections. In the legend, the original notation for each direction (symmetry group for forsterite, optical direction for enstatite) and its corresponding polarization with respect to the crystallographic axes (a,b,c) is noted, as well as the weight the direction would get assuming randomly oriented particles. Crystalline, directional crystals have different cross sections than those extensively utilized in exoplanet literature. Cross sections were computed using the Mie scattering algorithm adapted from LX-MIE \citep{KitzmannHeng2018} and described in \citet{Mullens2024}. Refractive index references follow from Table 1 in \citet{Mullens2024}: WS15 \citep{Wakeford2015}, KH18 \citep{KitzmannHeng2018}, B21 \citep{Burningham2021}, gcmCRT  \citep{Lee2022}. New references in this work are Z13 \citep{Zeidler2013}, E13 \citep{Eckes2013}, and Z15 \citep{Zeidler2015}, which are detailed in \S \ref{sec:cross_sections} and Table 1 in Zenodo Supplementary Material (see \S \ref{sec:links}).}    
    \label{fig:cross-sections-directionality}
\end{figure*}

\citet{Wakeford2015}, \citet{KitzmannHeng2018}, and \citet{Mullens2024} all compiled refractive indices for aerosols relevant to exoplanet studies. While \citet{Wakeford2015} does not explicitly explore the effect of aerosol orientation, \citet{KitzmannHeng2018} and \citet{Mullens2024} makes note of when the refractive index data for specific species (like graphite, quartz, ADP, etc.) are anisotropic. There have been a variety of methods to account for directionality in aerosol data: the so called spectral averaging method (SA) where cross sections are averaged \citep[][Section 9.3]{Bohren1983}, and the optical constant averaging method (OCA) where refractive indices or dielectric functions are averaged \citep{Mogli2007,Reed2017}. In \citet{KitzmannHeng2018} and \citet{Mullens2024}, the OCA method is used where the mean refractive indices for anisotropic species were computed by weighting the dielectric functions/refractive indices in each direction by 1/3. In \citet{Zeidler2013}, the SA method is used where cross sections are derived for each direction and then averaged. 

For our study, we analyze the direction-dependent refractive indices of three silicates that have been predicted and observed in sub-stellar atmospheres: quartz (SiO$_2$), forsterite (Mg$_2$SiO$_4$), and enstatite (MgSiO$_3$) \citep[e.g.,][]{Helling2006,Visscher2010,Lee2016}. Our quartz refractive indices come from \citet{Zeidler2013}\footnote{Sourced from the Database of Optical Constants for Cosmic Dust: \url{https://www2.astro.uni-jena.de/Laboratory/OCDB/crsilicates.html}} and \citet{DeSousaMeneses2014}. \citet{Zeidler2013} explore the temperature and directional dependence of quartz's spectral properties by taking IR reflection measurements of a natural $\alpha$ quartz crystal from Brazil. Their measurements range from 300~K to 928~K, capturing the $\alpha$ to $\beta$ quartz phase transition that occurs at 846-847~K. \citet{DeSousaMeneses2014} explore the temperature dependence of quartz's spectral properties by making spectral reflectivity and emissivity measurements of a cut, quartz plate, measuring only the E-symmetry (corresponding to the ordinary ray). Their measurements range from 295~K to 1880~K and capture the $\alpha$ to $\beta$ quartz phase transition at 846~K, and the $\beta$ quartz to $\beta$ cristobalite transition at 1743~K. As noted above, $\beta$ cristobalite is cubic and therefore isotropic, and therefore will not have the same directional dependence as $\alpha$ and $\beta$ quartz. 

Our forsterite refractive indices come from \citet{Eckes2013}. They explore the temperature and directional dependence of forsterite's spectral properties by taking polarized reflectance and emittance spectra of synthetic single crystals. Their measurements range from 295K to $\sim$1800-2000K. Using the D$_{2h}$ symmetry group, we define E$\parallel$z = B1U = E$\parallel$c, E$\parallel$y = B2U = E$\parallel$b, E$\parallel$x = B3U = E$\parallel$a (pers comm., D. D. S. Meneses).

Our enstatite refractive indices come from \citet{Zeidler2015}\footnote{Sourced from the Database of Optical Constants for Cosmic Dust: \url{https://www2.astro.uni-jena.de/Laboratory/OCDB/crsilicates.html}}. They explore the temperature and directional dependence of enstatite's spectral properties by taking IR reflection measurements of a natural, single crystal from Burma. Their measurements range from 10~K to 928~K. The enstatite used in their study is specifically orthoenstatite. Pyroxenes like enstatite can form with a monoclinic (clinopyroxene) or orthorhombic (orthopyroxene) crystal shape. Both crystal shapes result in a biaxial crystal, however orthopyroxene is more common in nature \citep{Jaeger1998}. \citet{Zeidler2015} determines that their orthoenstatite has some Fe inclusions through EDX analysis, resulting in a more accurate chemical formula of Mg$_{0.92}$Fe$_{0.09}$SiO$_3$. We define E$\parallel$x=E$\parallel$b, E$\parallel$y=E$\parallel$a, E$\parallel$z=E$\parallel$c as defined on the Database of Optical Constants for Cosmic Dust\footnote{\url{https://www2.astro.uni-jena.de/Laboratory/OCDB/crsilicates.html}}.

Additional temperature and directional dependent refractive indices for aerosols relevant to exoplanets are discussed in more detail in \S \ref{sec:aerosol_database}. Effective extinction cross sections ($\sigma_{\mathrm{ext,eff}}$) for each species are computed via the algorithm discussed in \citet{Mullens2024}, assuming spherical particles. We explore both the optical constant averaging method (OCA) by averaging refractive indices before computing the cross section, and the spectral averaging method (SA) where cross sections are computed from each directional refractive index and then averaged. We compare the resultant directional $\sigma_{\mathrm{ext,eff}}$ for 0.01 \textmu m sized particles (with a 0.5 width lognormal distribution) with commonly used refractive indices used in exoplanet literature in Figure \ref{fig:cross-sections-directionality}. \nt{We specifically compare room temperature directional cross sections with cross sections commonly used in exoplanet literature (which nominally do not account for temperature or specific directionality explicitly) in order to showcase the specific effects of directionality on resultant cross sections. Additionally, room temperature measurements are the most ubiquitously measured and validated in laboratory studies, with temperature dependent measurements being a field of active development \citep{Eckes2013,DeSousaMeneses2014,Zeidler2015}}.We compare the OCA and SA cross section averaging methods in Figure~\ref{fig:cross-sections-averages}. We over-plot directional dependent cross sections of room temperature and `hot' ($\sim$ 1000K) silicate aerosols in Figure~\ref{fig:cross-sections-temperature}. 

Crystalline quartz has the least extreme directional dependence in the 8-9 \textmu m region, with ordinary and extraordinary cross sections having absorption features near the same wavelength region (Figure \ref{fig:cross-sections-directionality}, top). The ordinary cross section (aqua) compared to the extraordinary cross section (green) for room temperature $\alpha$ quartz displays a double-peaked absorption feature at 8-9 \textmu m, and an absorption feature shifted red-ward in the 12-13 \textmu m region. At hotter temperatures (Figure \ref{fig:cross-sections-temperature}, top), $\beta$ quartz's ordinary cross section (orange) double-peaked feature becomes smoothed out and the extraordinary cross section (magenta) no longer has any absorption in the 12-13 \textmu m region. At the highest temperatures, $\beta$ cristobalite (red) is no longer directional dependent and has its absorption feature shifted redward. When compared to archival cross sections, the resultant absorption features agree in location (but not morphology) at the 8-9 \textmu m region, and differ in the 12-13 \textmu m region.

Crystalline forsterite and enstatite have more extreme directional dependence. Room temperature forsterite has three distinct, sharply peaked absorption features in the 9.5 - 11.25 \textmu m region (Figure \ref{fig:cross-sections-directionality}, middle, from shortest to longest wavelength: E$\parallel$a (blue), E$\parallel$c (green), E$\parallel$b (turquoise)) that all shift to longer wavelengths with increasing temperature (Figure \ref{fig:cross-sections-temperature}, middle). The optical properties derived from popular refractive indices used in exoplanet literature seems to match one or two of these features, but not all three. Room temperature enstatite has two sharply peaked absorption features near the same wavelength in the 8.5 - 9.5 \textmu m region (Figure \ref{fig:cross-sections-directionality}, bottom, E$\parallel$b (turquoise), E$\parallel$a (blue)) with all three directions having overlapping cross sections in the 10 - 12 \textmu m region (where E$\parallel$c (green) has the strongest cross section at 10.2 \textmu m). Cross sections once again shift to longer wavelengths with increasing temperature (Figure \ref{fig:cross-sections-temperature}, bottom). 

As seen in Figure \ref{fig:cross-sections-averages}, we find that averaging refractive indices (OCA, dotted line) before computing cross sections can lead to drastic differences in resultant cross sections, especially in the case of forsterite (Mg$_2$SiO$_4$, middle). In particular, the three directional absorption features for the averaged cross section of Mg$_2$SiO$_4$ occur around 9.7 \textmu m, 10.6 \textmu m, and 10.8 \textmu m while the averaged refractive index cross section has a single absorption feature around 10.1 \textmu m.

\section{\nt{Model Transmission and Emission Spectra}}\label{sec:model-spectra}

In order to test our observational sensitivity to directional cross sections of aerosols, we generate model transmission and emission for the hot Jupiter WASP-17b in the wavelength range of JWST-MIRI-LRS (5 - 14 \textmu m). We use the room temperature effective extinction cross sections from \S \ref{sec:cross_sections} since room temperature cross sections are the most ubiquitously measured and validated, and defer temperature dependence to future work \citep[e.g.,][]{Moran2024}.

We generate the forward model spectra with the open-source atmospheric retrieval code \texttt{POSEIDON}\footnote{\url{https://github.com/MartianColonist/POSEIDON}} \citep{MacDonaldMadhusudhan2017, MacDonald2023, Mullens2024}. In transmission and emission, we assume a one-dimensional H$_2$-He dominated atmosphere (with He/H$_2$ = 0.17) with water in the gas-phase. Here we assume a simplified hot Jupiter atmosphere with gas-phase opacity primarily from continuum opacity from H$_2$ and He collision-induced absorption \citep{Karman2019} and H$_2$O absorption \citep{Polyansky2018}, with the inclusion of aerosol absorption (as described in \citet{Mullens2024}). 
The reference pressure was set at 10\,bar with model atmospheres covering 10$^{-8}$ - 100 bar (10$^{-6}$ - 100 bar for emission) with 100 layers uniformly distributed in log-pressure space. For transmission we assume an isothermal pressure-temperature profile and the retrieved properties from analysis of the full data-set (archival HST + Spitzer, JWST NIRISS SOSS, and JWST MIRI LRS) in \citet{Louie2024}\footnote{R$_{\mathrm{p,ref}}$ = 1.68 R$_\mathrm{J}$, T = 1271.9 K, log H$_2$O = -2.96, log P$_\mathrm{top,slab}$ = -6.60, $\Delta$ log P = 1.96, log r$_{\mathrm{m}}$ = -1.85, log SiO$_2$/Mg$_2$SiO$_4$/MgSiO$_3$ = -11.23} which indicate a thin slab composed of sub-micron sized particles in the atmosphere. For emission we assume the `Guillot' profile \citep{Guillot2010} as described in \citet{Mullens2024} with pressure-temperature and aerosol properties chosen to best demonstrate the absorption features (a nearly uniform in the atmosphere aerosol distribution) due to the JWST-MIRI-LRS emission data having a non-detection of dayside aerosols \citep{Valentine2024}\footnote{R$_{\mathrm{p,ref}}$ = 1.68 R$_\mathrm{J}$, log $\kappa_{\mathrm{IR}}$ = -4, log $\gamma$ = -0.4, T$_{\mathrm{int}}$ = 600, T$_{\mathrm{equ}}$ = 1490, log H$_2$O = -5, log P$_\mathrm{top,slab}$ = -6, $\Delta$ log P = 8, log r$_{\mathrm{m}}$ = -2, log SiO$_2$/Mg$_2$SiO$_4$/MgSiO$_3$ = -13}. In emission, we do not consider multiple scattering in our models since the particle size (sub-micron) is much smaller than the wavelength of the spectra, and defer to future work to explore the scattering effects of $\sim$micron-sized particles that could be relevant for high-mass objects such as brown dwarfs.

\begin{figure*}
    \includegraphics[width=1.0\textwidth]{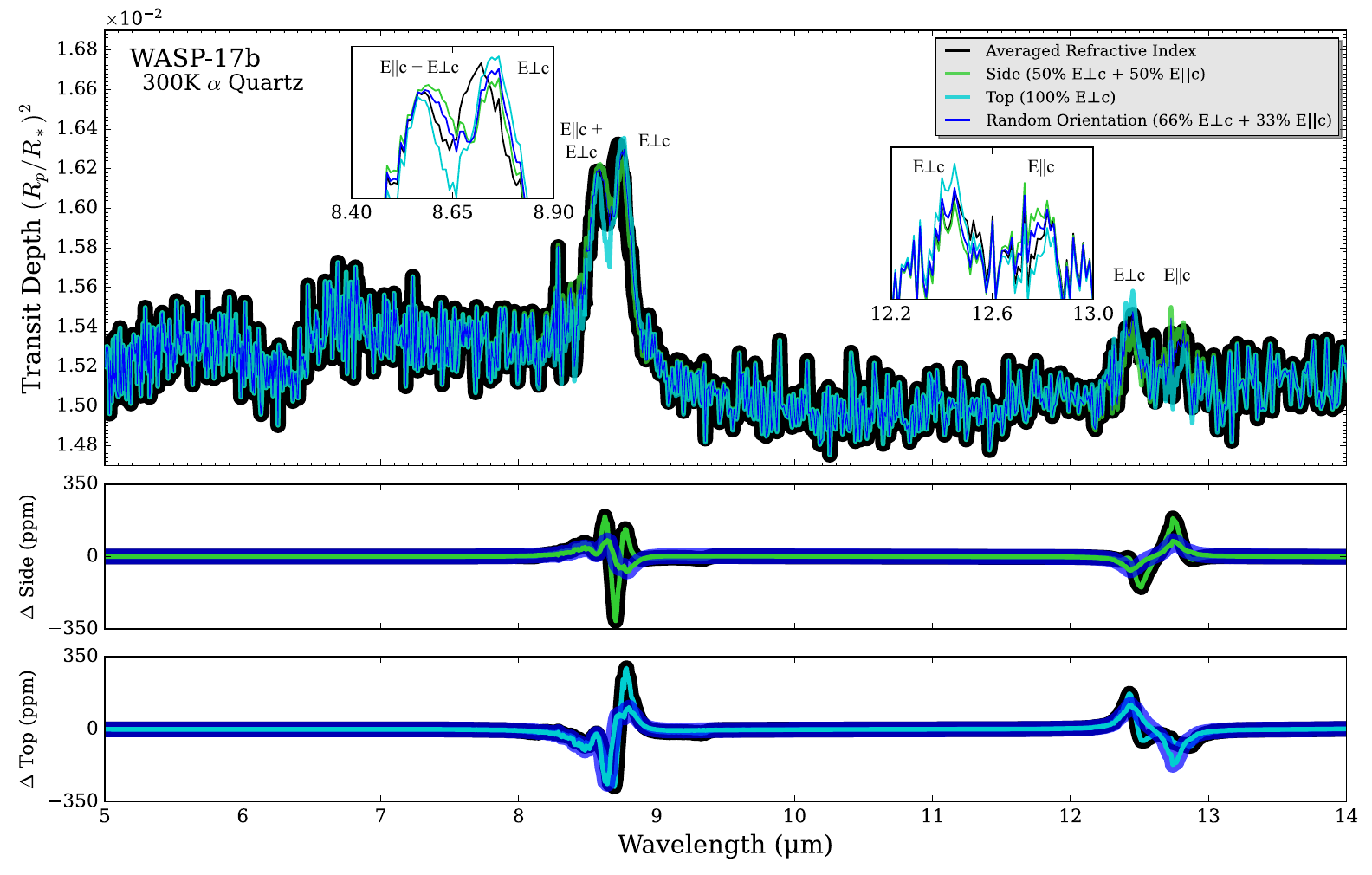}
    \caption{Example mid-infrared transmission spectra with directional quartz (SiO$_2$) of the hot Jupiter WASP-17b utilizing the retrieved atmospheric and aerosol parameters from \citet{Louie2024}. \textbf{Top}: Two spectra were generated assuming randomly oriented aerosols: averaged refractive index (OCA, black) and averaged cross section (SA, blue) [See Figure \ref{fig:cross-sections-averages}]. Two spectra were generated assuming oriented aerosols: Top (expected orientation in transmission assuming mechanical alignment, green) and Side (expected orientation in emission assuming mechanical alignment, cyan) [See Figure \ref{fig:crystal-art}]. Insets zoom-in on regions of aerosol absorption. \textbf{Bottom}: PPM differences between oriented and randomly oriented spectra: side - averaged refractive index (green + black outline) and side - random orientation (green + blue outline), with a similar procedure done with top. Quartz's directionality is determined by the morphology of the absorption feature around $\sim$ 8.5 \textmu m. While it is difficult to distinguish Side vs random orientation for quartz crystals, Top can be distinguished from random orientation within $\sim$ 200 ppm. 
    Spectra were computed with \texttt{POSEIDON} \citep{MacDonaldMadhusudhan2017,MacDonald2023,Mullens2024}. [See Figure \ref{fig:sio2_emission} for similar plot, but in emission.]}    
    \label{fig:sio2_transmission}
\end{figure*}

We model the directionality of quartz four ways, where results for transmission can be seen in Figure \ref{fig:sio2_transmission} and emission in \ref{fig:sio2_emission}. The first is using the cross section computed by averaging the refractive index (OCA, black). We then model the grains assuming that the grains are mechanically aligned with the wind\footnote{See \S \ref{POSEIDON_updates} for how the percentages are applied to the bulk aerosol mixing ratio}; propagation parallel to c (Top, green) corresponds to the upper right panel of Figure \ref{fig:crystal-art} and is what we expect for transmission geometries while propagation perpendicular to c (Side, turquoise) corresponds to the upper left panel and is what we expect for dayside emission geometries. Finally we assume that grains are not mechanically aligned with the wind and instead have a random orientation (SA, blue). In the bottom panels, we show the ppm difference between the averaged refractive index (black outline)/random orientation (blue outline) and the mechanically aligned spectra. 

\begin{figure*}
    \includegraphics[width=1.0\textwidth]{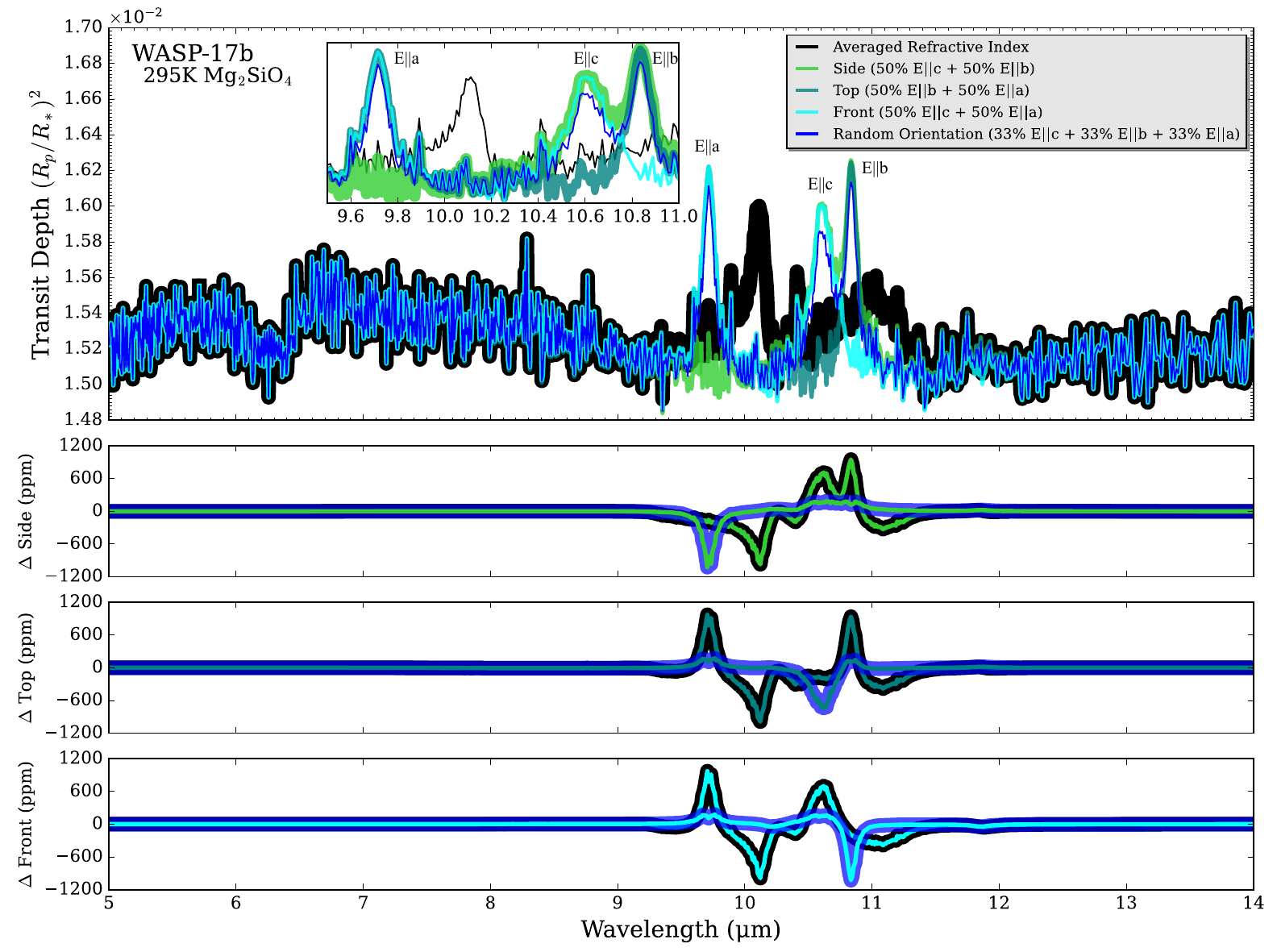}
    \caption{Same as Figure \ref{fig:sio2_transmission}, but with forsterite (Mg$_2$SiO$_4$). In contrast to quartz, the three distinct absorption features (9.7, 10.6, and 10.9 \textmu m) of crystalline forsterite, in lieu of absorption peak morphology, are diagnostic of orientation.
    The Top orientation (expected orientation in transmission assuming mechanical alignment, turquoise) only has absorption features at $\sim$ 9.7 and 10.9 \textmu m (missing the feature at 10.6 \textmu m). The Side and Front orientations (expected orientation in emission assuming mechanical alignment, green and aqua) also only have two out of three absorption features. All three configurations have $\sim$ 1000 ppm differences between random orientation (assuming Random Orientation, SA (blue) to be preferred) and orientated grains. [See Figure \ref{fig:mg2sio4_emission} for similar plot, but in emission].}    
    \label{fig:mg2sio4_transmission}
\end{figure*}

\begin{figure*}
    \includegraphics[width=1.0\textwidth]{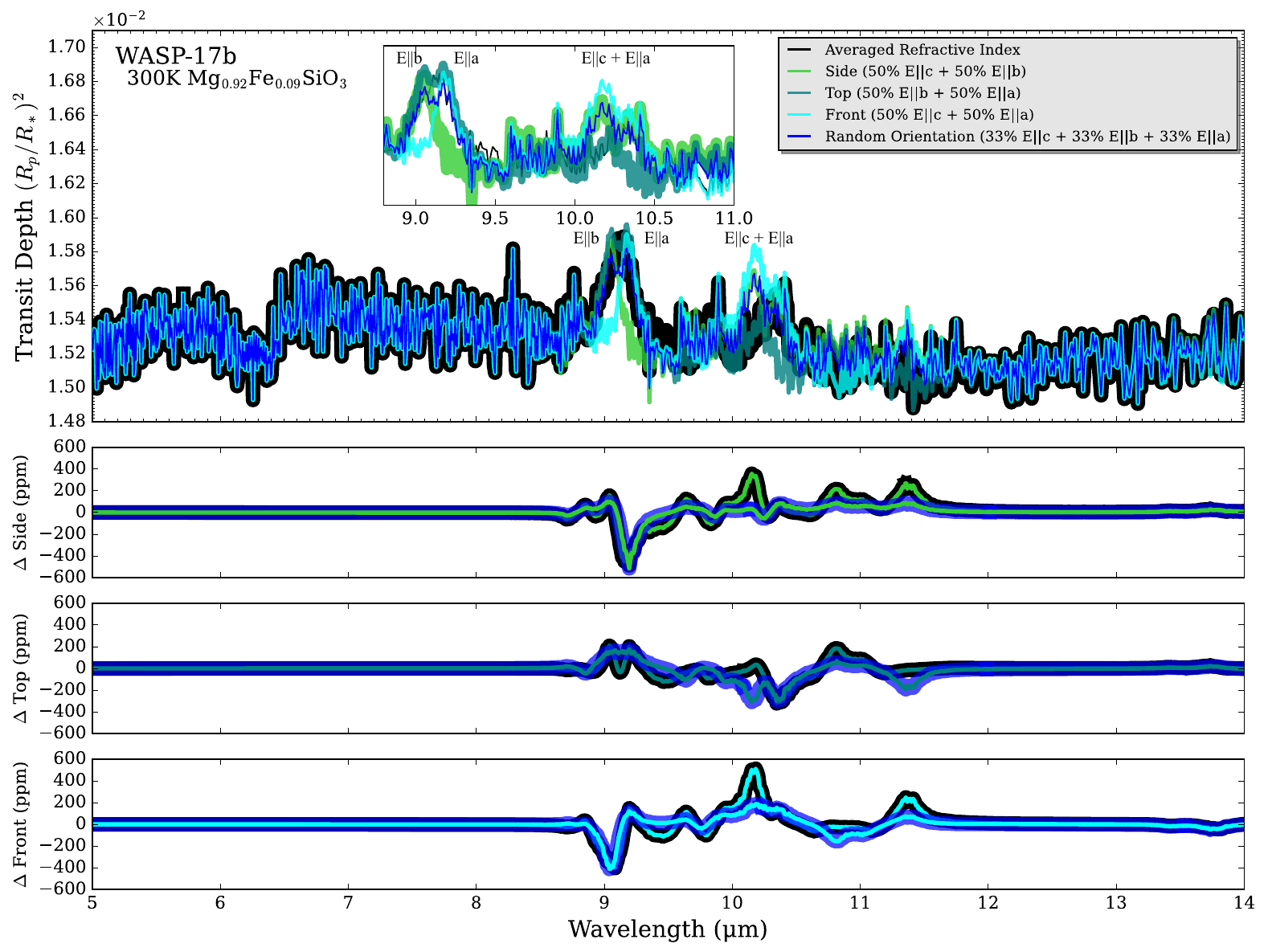}
    \caption{Same as Figure \ref{fig:sio2_transmission} and \ref{fig:mg2sio4_transmission}, but with (ortho-)enstatite (Mg$_{0.92}$Fe$_{0.09}$SiO$_3$). Enstatite has directionality determined by both absorption feature morphology and peak position. The Top orientation (expected orientation in transmission assuming mechanical alignment, turquoise) would have a broadened absorption feature at $\sim$ 9.2 \textmu m due to the combined E$\parallel$b and E$\parallel$a cross sections, but not the E$\parallel$c feature. The Side and Front orientations (expected orientation in emission assuming mechanical alignment, green and aqua) have a less broad absorption feature at $\sim$ 9.2 \textmu m that is shifted to the left or right (respectively) in wavelength space. 
    Side and Front have $\sim$ 1000 ppm differences between random orientation (assuming Random Orientation, SA (blue) to be preferred) and orientated grains, while `Top' has a maximum $\sim$ 200 ppm difference. [See Figure \ref{fig:mgsio3_emission} for similar plot, but in emission].}    
    \label{fig:mgsio3_transmission}
\end{figure*}

For forsterite and enstatite, results for forsterite in transmission can be seen in can be seen in Figure \ref{fig:mg2sio4_transmission} and emission in \ref{fig:mg2sio4_emission} and for enstatite in transmission in Figure \ref{fig:mgsio3_transmission} and emission in \ref{fig:mgsio3_emission}. We similarly model the cross section computed by averaging the refractive index (OCA, black) and assuming the grains are not mechanically aligned with the wind and have random orientation (SA, blue). Since these crystals are biaxial, we have to model how they are mechanically aligned with the wind differently than quartz: `Top' (dark turquoise) assumes that light is propagating parallel to the c-axis and is what we expect for transmission geometries while `Side' (green) and `Front' (aqua) assume that light is propagating parallel to the a and b axis (respectably) which can be expected for dayside emission geometries. Similar to quartz, the ppm difference in spectra are displayed in the bottom panels. 

We find $\geq$ 100 ppm differences between spectra can arise from using the cross section derived from averaging refractive indices (OCA) vs weighing the cross sections (SA) (ex., see ppm differences w/ black outline in bottom panels of Figure \ref{fig:mg2sio4_transmission}), displaying that care must be given when modeling crystalline aerosol cross sections. The detectability of directionality with different aerosols arises in two ways: the first is absorption feature morphology, which is prevalent for quartz, and the second is absorption feature peak position, which is prevalent for forsterite. 

The orientation of quartz grains is detectable ($\geq$ 100 ppm difference when compared to random orientation) when light is propagating parallel to the c-axis, which is the orientation we expect to observe if grains are mechanically aligned with the gas flow on the limb. Light propagating perpendicular to the c-axis, which is the orientation we expect to observe if grains are mechanically aligned on the dayside, is more difficult to distinguish from randomly oriented grains. This is due to the double-peaked absorption feature in the ordinary cross section being smoothed out by the extraordinary cross section. Therefore, the detection of directionality of quartz is strongly dependent on the absorption feature morphology itself.

We find that forsterite has the strongest signature of directionality of the silicates we explore due to having its directionality resulting in three distinct absorption peak positions. In particular, we find that each direction (`Side', `Top', and `Front') only has two of the three main absorption features. In transmission, mechanically aligned forsterite (`Top') would only have absorption features at $\sim$ 9.7 and 10.9 \textmu m and lack the feature at 10.6 \textmu m. In emission, both the `Side' and `Front' configurations are permissible. If one of the two dayside configurations are more dynamically favorable than the other, it would be detectable via a similar two out of three feature approach. If both are favorable, all three absorption features would be present in the data but the relative strength of the E$\parallel$c feature at 10.6 \textmu m would be stronger than the other two (since both `Side' and `Front' have contributions from E$\parallel$c indices). The `Side' vs `Front' vs `Side + Front' absorption features comparison is displayed in Figure \ref{fig:side-front}.

Enstatite represents a mid-case between quartz and estatite since its directionality is determined by absorption feature morphology and peak position. Two of its features (E$\parallel$b at 9.0 \textmu m and E$\parallel$a at 9.3 \textmu m) are distinct but overlap substantially while the third feature (E$\parallel$c at 10.2 \textmu m) is mostly distinct (with some contributions from all three polarizations). In transmission, mechanically aligned enstatite (`Top') would only have a broadened absorption feature at $\sim$ 9.2 \textmu m due to the combined E$\parallel$b and E$\parallel$a cross sections, but not the E$\parallel$c feature. In emission geometries, if the `Side' or `Front' configurations are favorable, the absorption feature will be thinner and shifted to the left or right (respectively) in wavelength space. Otherwise, if both are favorable the feature would be broadened but have a shorter amplitude while the E$\parallel$c at 10.2 \textmu m would be stronger. The `Side' vs `Front' vs `Side + Front' absorption features comparison is displayed in Figure \ref{fig:side-front}.

\begin{figure*}
    \includegraphics[width=1.0\textwidth]{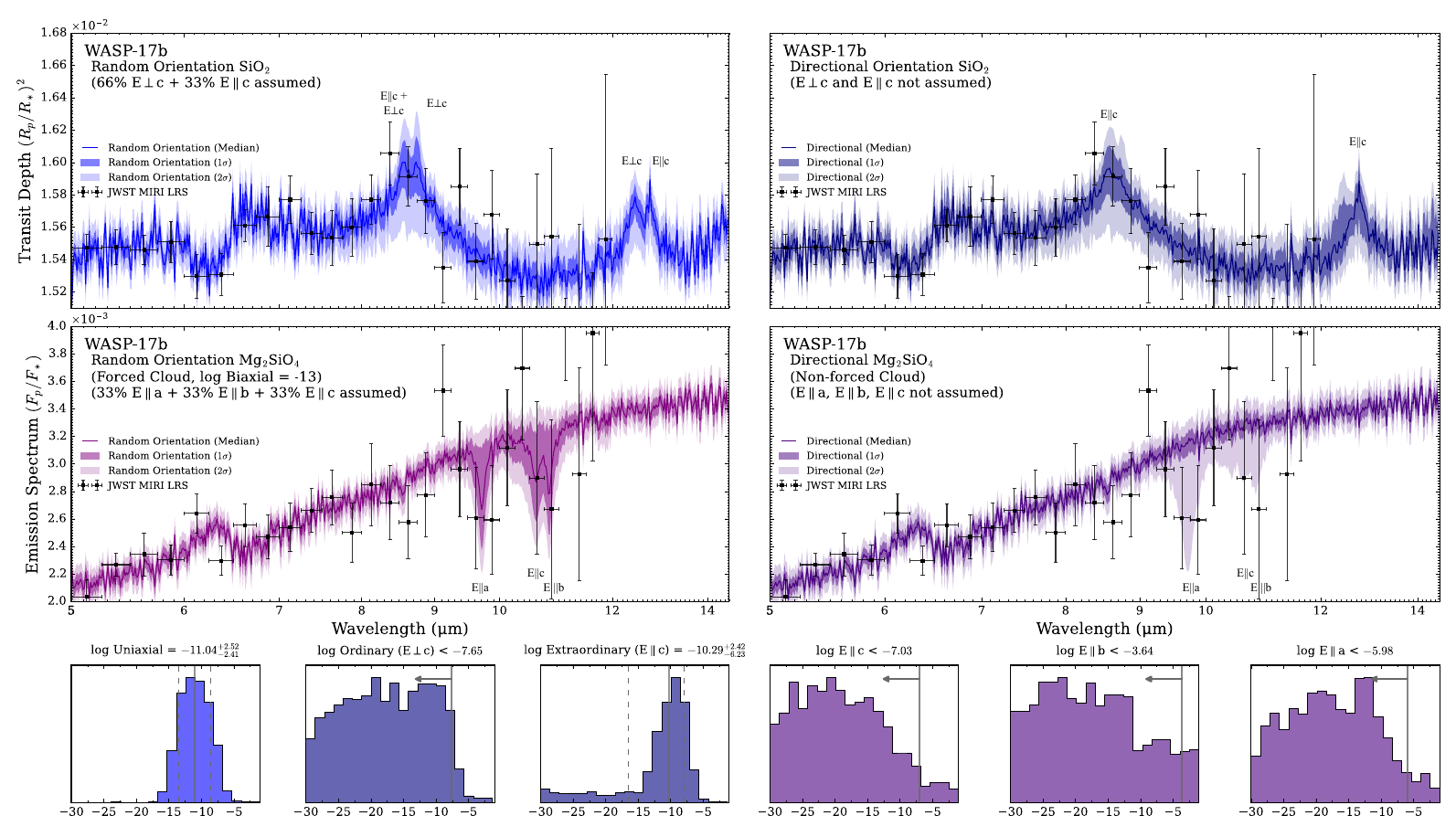}
    \caption{Atmospheric retrieval of JWST-MIRI-LRS transmission and emission spectrum of WASP-17b \citep{Grant2023,Valentine2024}, highlighting the four new aerosol models (unaixial and biaxial, random orientation vs directional) in \texttt{POSEIDON}. See \S \ref{POSEIDON_updates} for details on priors and model set-up. (Top, Left) Random orientation SiO$_2$ assuming a 1/3 extraordinary + 2/3 ordinary weight on the mixing ratio. (Top, Right) Directional SiO$_2$ which retrieved the extraordinary and ordinary mixing ratios separately. \nt{The retrieval prefers the extraordinary cross section over the ordinary one, which is nonphysical (as the extraordinary cross section can only reach a maximum of 50\% weight).} 
    (Bottom, Left) Random orientation Mg$_2$SiO$_4$ assuming a 1/3 weight for all three directional cross section mixing ratios E $\parallel$ c, b, a where we `forced' the cloud with strict priors on aerosol properties to explore parameter space that favored a cloud. (Bottom, Right) Directional Mg$_2$SiO$_4$ which retrieved the E $\parallel$ c, b, a mixing ratios separately, where we utilized uniform uninformative priors. (Bottom Row) Retrieval posteriors for aerosol mixing ratios, with colors corresponding to to the retrievals. Discussion of retrieval results can be found in \S \ref{POSEIDON_updates}
    } 
    \label{fig:directionality-retrievals}
\end{figure*}



\section{Retrieval Tests on WASP-17b}

Here we expand the retrieval capabilities of the open-source code \texttt{POSEIDON} to consider aerosol directionality, and test the code on JWST MIRI-LRS transmission and emission data of the hot Jupiter WASP-17b, which was determined to have crystalline quartz crystals in its atmosphere in \citet{Grant2023}.

\subsection{\texttt{POSEIDON} Updates and Retrievals}\label{POSEIDON_updates}

We have updated the pre-computed aerosol database in the open-source atmospheric retrieval code \texttt{POSEIDON}\footnote{\url{https://github.com/MartianColonist/POSEIDON}} \citep{MacDonaldMadhusudhan2017, MacDonald2023} presented in \citet{Mullens2024} with directional and temperature dependent aerosol cross sections. The updates to the database are detailed further in \S \ref{sec:aerosol_database}. In addition, we have included four new aerosol models to the code to take into account aerosol directionality. 

The four new aerosol models are an extension of the `slab' model found in \citet{Mullens2024}. Briefly, the slab aerosol model is described by a slab top pressure, $P_{top, \, slab}$, a vertical slab width, $\Delta P$, the (constant) aerosol mixing ratio within the slab, and the mean particle radius, $r_m$. 

The first variations are the `uniaxial random slab' and `biaxial random slab' aerosol models. These model's free parameters are the same as the nominal slab model and assume that the aerosols in the atmosphere are randomly oriented (i.e. not mechanically oriented). These models fit for a single aerosol mixing ratio (log X$_\mathrm{uniaxial}$ and log X$_\mathrm{biaxial}$)\footnote{\nt{All logs in this paper are log base 10.}} and subsequently assume that the mixing ratio of directional cross sections are 
\begin{align}
    \mathrm{log_{\mathrm{10}}} \; \mathrm{X_{ordinary}} &= \mathrm{log}_{10}(2/3*10^{\mathrm{log_{\mathrm{10}}} \; \mathrm{X_{uniaxial}}})\\
    \mathrm{log_{\mathrm{10}}} \; \mathrm{X_{extraordinary}} &= \mathrm{log}_{10}(1/3*10^{\mathrm{log_{\mathrm{10}}} \; \mathrm{X_{uniaxial}}}) \\
    \mathrm{log_{\mathrm{10}}} \; \mathrm{X_{E\parallel c,b,c}} &= \mathrm{log}_{10}(1/3*10^{\mathrm{log_{\mathrm{10}}} \; \mathrm{X_{biaxial}}})
\end{align}
with the same mean particle radius assumed for all aerosols. Applying the weights to the mixing ratios is equivalent to applying them to the cross sections since \texttt{POSEIDON} computes the extinction coefficient ($\kappa$) as 
\begin{align}
\kappa = \mathrm{sum}(n_{\mathrm{tot}} \; \times \; X_{i} \; \times \; \sigma_{\mathrm{ext,eff}})
\end{align}
where $n_{\mathrm{tot}}$ is the pressure-dependent total atmospheric number density and $X_{i}$ is the mixing ratio. Spectral features do not scale linearly with the weight on the cross section/mixing ratio, since the attenuation and resultant spectrum scales with e$^{-\kappa \; ds}$, which is non-linear relationship. See \citet{MacDonald2023} for more details on how attenuation is computed in \texttt{POSEIDON}. The other variations of the slab model are `uniaxial slab' and `biaxial slab' that instead have the mixing ratio for each directional cross section be an independent free parameter (while still assuming the same mean particle radius), which can be used to retrieve oriented aerosols\footnote{For a tutorial on how to implement directionality aerosol models in \texttt{POSEIDON}, see the `Aerosol Advanced: Directionality’ tutorial in \href{https://poseidon-retrievals.readthedocs.io/en/latest/content/forward_model_tutorials.html}{Forward Model Tutorials}}. 

We ran retrievals using these models on the JWST MIRI LRS transmission and emission data of WASP-17b. We are limited in only retrieving on JWST MIRI LRS data due to our refractive indices for directional aerosols only having coverage in the mid-infrared (5-30 \textmu m). 
We chose to set informative priors on our retrieval exploration based on extant retrieval analysis of WASP-17b. The transmission data has a clear SiO$_2$ feature present, therefore for transmission we set informative priors based on the retrieved isothermal temperature and water from \citet{Grant2023} and \citet{Louie2024} and set uninformative wide priors on aerosol properties (in both the random and directional orientation retrievals). In \citet{Valentine2024}, the precision of the emission dataset did not allow for a conclusive detection of aerosol absorption, but a Gaussian significance test on the 9-12 \textmu m region resulted in a 2.5-2.8 $\sigma$ detection significance that the unexplained features were of astrophysical origin that could be possibly explained via more complex cloud analysis. For this reason, we set fairly tight priors on the PT profile and H$_2$O abundances for all emission retrievals, and for the random orientation retrievals we set a tight priors on the aerosol abundance to explore parameter space that favors the inclusion of dayside clouds. For the directional retrieval, we loosened this tight aerosol abundance prior and allowed the retrieval to explore both clear and cloudy solutions. All priors and free parameters for each retrieval are highlighted in Table 2 in the Zenodo Supplementary Material (see \S \ref{sec:links}). For transmission we tested SiO$_2$ while for emission we tested both SiO$_2$ and Mg$_2$SiO$_4$. Results for a subset of retrievals are highlighted in Figure \ref{fig:directionality-retrievals}, with the full set of retrievals presented in Zenodo Supplementary Material Figure 2 (see \S \ref{sec:links}).

We find that, in transmission, oriented crystals are preferred over random orientation by $\sim$ 1.4 $\sigma$, but the extraordinary cross sections are the preferred solution due to the leftward shift found in the absorption peak of the data (in lieu of the ordinary indices, which feature a double-peaked feature). This result is highly nonphysical, as the highest the extraordinary absorption can go is 50\%, as discussed in \S \ref{sec:cross_sections}. We discuss potential properties of aerosols that can result in a shifted feature in \S \ref{sec:discussion}. 

In emission, there are two deviations in the dataset explored in \citet{Valentine2024} as possible aerosol absorption features. In our `forced cloud' retrieval we do find that crystalline Mg$_2$SiO$_4$ can explain potential features there, but in the directional retrieval without tight priors on the aerosol abundance, these features are only fit in the 2$\sigma$ shaded region. 
\nt{In order to statistically quantify any preference of directionality over random orientation, we performed four additional retrievals: a random orientation retrieval, and three retrievals with only one directional cross section (ex. E$\parallel$c alone). These retrievals were configured with the same exact priors (as detailed in Table 2 in the Zenodo Supplementary Material, see \S \ref{sec:links}) and number of free parameters. We find that E$\parallel$a alone is preferred over randomly-oriented particles by 1$\sigma$, while it is preferred over both E$\parallel$b alone and E$\parallel$c alone by 1.3$\sigma$. This could hint at either the `Top' or `Front' configurations highlighted in Figure \ref{fig:mg2sio4_emission}, which both contain the E$\parallel$a cross section. We note, however, that this slight statistical preference is perhaps due to the E$\parallel$a absorption feature (9.7 \textmu m) being in a region of higher SNR than E$\parallel$b (10.9 \textmu m) and E$\parallel$c (10.6 \textmu m).} Results for SiO$_2$ retrievals on the emission dataset are featured in Zenodo Supplementary Material Figure 2 (see \S \ref{sec:links}). This retrieval exploration displays the potential power of MIRI-LRS at detecting grain directionality and the interaction of grains with local dynamics in Hot Jupiter atmospheres.

\subsection{\nt{Observing Strategies}}

\nt{Devising an optimal JWST observation strategy to detect aerosol orientation is dependent on your proposed target. JWST MIRI LRS has been successful in detecting silicate absorption features in transmission \citep{Grant2023,Dyrek2024} and emission \citep[e.g.,][]{Ingles2024} spectroscopy, while JWST MIRI MRS has been successful in detecting silicates in directly imaging spectra \citep[][]{Miles2023}, though JWST MIRI MRS requires sufficiently bright targets to do so.} 

\nt{The directionality of quartz is highly localized in wavelength space (see Figure \ref{fig:sio2_transmission}), and therefore will require a high enough spectral resolution in the 8-10 \textmu m region to differentiate. The directionality of forsterite is dependent on three distinct absorption features (see Figure \ref{fig:mg2sio4_transmission}), and therefore depends on the detecting absorption features in the 9.5-11 \textmu m region (which is typically a region of lower SNR due to MIRI LRS's throughput).}

\nt{In order to observe aerosol directionality, which provides a powerful tool to probe aerosol-gas coupling as well as other second-order effects (see \S \ref{sec:discussion}), we recommend exploring targets with confirmed silicate detections (e.g., WASP-17b, WASP-107b, HD-189733b, VHS 1256b) and different observing strategies. Observing strategies for JWST MIRI LRS involve stacking transit and eclipse observations, or saturating the blue-edge of the detector to obtain higher SNR in wavelengths where silicates have absorption features. Both of these strategies will be explored in an upcoming JWST Cycle 4 program (JWST-GO-08309, PIs: Mullens \& Moran). JWST MIRI MRS requires bright, low-gravity targets. For generalized precisions necessary to detect aerosol directionality, see the ppm difference plots present in Figures \ref{fig:sio2_transmission}, \ref{fig:mg2sio4_transmission}, \ref{fig:mgsio3_transmission},
\ref{fig:sio2_emission}, \ref{fig:mg2sio4_emission}, and \ref{fig:mgsio3_emission}.}

\section{Discussion and Conclusions}\label{sec:discussion}

In \S \ref{sec:crystalline-aerosol-anisotropy} we assert that aerosols in hot Jupiter atmospheres are expected to be crystalline, either via direct condensation or annealing, which has been supported by previous JWST observations of WASP-17b \citep{Grant2023}. We note, however, that the silicate feature of WASP-107b utilized a mixture of amorphous  SiO, SiO$_2$, MgSiO$_3$, and C to obtain a good fit to their data \citep{Dyrek2024}. Silicate clouds in hot Jupiter atmospheres are thought to directly nucleate from the reaction of vapor phase oxygen and silicon, and can form in the amorphous state at lower temperatures \citep{Fabian2000}. This can occur globally in hot Jupiters with lower equilibrium temperatures (WASP-107b T$_{\mathrm{equ}}$ = 770K vs. WASP-17b  T$_{\mathrm{equ}}$ = 1750K), or potentially on the cooler nightsides of highly irradiated hot Jupiters (where aerosols can then get advected to the western limb). Crystalline aerosols can form directly from the vapor phase in hot environs (the dayside and eastern-limb of highly irradiated hot Jupiters), or from the annealing of amorphous aerosols (which occurs on the timescales of hours) due to transport of amorphous particles to hotter environs. The detected phase of an aerosol (amorphous vs crystalline) probes where aerosols are forming and the global cycling of aerosols in the atmosphere; if amorphous aerosols form on the nightside and are advected to the dayside, they will anneal if the temperatures are high enough to form crystalline aerosols, assuming they aren't sublimated back to vapor form. If they don't sublimate, the predicted stable phase of aerosols in the global atmosphere will be crystalline \ntt{(i.e., all amorphous particles will anneal and over global timescales be predominately crystalline).  If crystalline quartz is directly condensing from the gas-phase on the eastern limb, there will be no amorphous aerosols in the atmosphere. The predicted global stable phase of aerosols in the atmosphere will be amorphous if amorphous particles are cycled to the dayside and do not get annealed. By probing for limb asymmetries \citep[e.g.,][]{Espinoza2024} with transmission spectroscopy, as well as probing the day and night emergent flux via emission spectroscopy and phase curves, the full story of aerosol formation and cycling can be uncovered. }

\nt{In \S \ref{sec:mechanical-alignment} and \S \ref{sec:aerosol-shape-appendix} we show that aerosols in hot Jupiter environments are expected to be non-spherical, meaning they are mechanically aligned with the wind. Both crystalline and amorphous aerosols can be non-spherical and therefore mechanically aligned, but this effect can only be probed with spectral observations via non-isotropic attenuation by crystalline aerosols. Additionally, crystalline aerosols can form sphere-like shapes if subjected to melting \citep{Kohout2014}. Sphere-shaped crystalline aerosols still have orientation-dependent attenuation will not be aligned with the wind, and therefore will have an opacity identical to the randomly oriented particles features in Figures \ref{fig:sio2_transmission}, \ref{fig:mg2sio4_transmission}, and \ref{fig:mgsio3_transmission}.}

In our work, we assume Mie theory and a tight lognormal particle radius distribution when modeling the optical properties of aerosols for simplicity. In future work we will consider more complex aerosol shapes, which we detail in \S \ref{sec:aerosol-shape-appendix}. \nt{We note, however, that assuming spherical particles and Mie theory has been sufficient when comparing to lab-measured opacities, even when considering aerosol directionality \citep[e.g.,][]{Zeidler2013}.}

In this work we explored two different ways to average the refractive indices of directional aerosols: the first being the so called spectral averaging method (SA) where cross sections are averaged \citep[][Section 9.3]{Bohren1983}, and the optical constant averaging method (OCA) where refractive indices (or dielectric functions) are averaged \citep{Mogli2007,Reed2017} before the cross section is computed. We find that the OCA method can cause large deviations and shifts in absorption feature location; in particular the three distinct absorption features of forsterite becomes a single absorption feature at a different wavelength (Figure \ref{fig:mg2sio4_transmission}). 

There are other methods by which to model both the directionality of refractive indices, as well as heterogeneous aerosols composed of more than one material. These so-called effective medium approximations (EMAs; Lorentz-Lorenz, Maxwell-Garnett, and Bruggeman) can be used to determine the optical properties and porosity of mixed materials and have been used successfully to mix clouds in exoplanet data analysis \citep[e.g.,][]{Dyrek2024} and forward modeling \citep{Kiefer2024} but can still have difficulties matching lab data \citep{Mill2019}.

Randomly oriented particles can also be simulated in the lab instead of post-processing of optical properties; \citet{Herbin2023} measure the optical properties of wind-borne crystalline quartz and used a polarizer to ensure that their particles were randomly oriented. Without a general consensus on best practices when it comes to modeling randomly oriented and oriented aerosols, it is important to consider both possibilities when fitting data.  

Our results are dependent on the lab data we use to compute the optical properties of the aerosols. Measured absorption and scattering properties of an aerosol sample depend on laboratory methodology (thin film vs KBr pellets vs cut crystals, vacuum vs air measurements), and whether a sample is natural or synthetic (natural crystals can have iron inclusions, which can enhance optical anisotropy in the presence of magnetism). For transparency, we have included these details in a Zenodo Supplementary Material Table 1 (see \S \ref{sec:links}). 

In \S \ref{sec:model-spectra} we discuss which absorption signatures we expect to see if grains are mechanically aligned with winds on the dayside or limb. There have been studies that explore the orientation of grains on the dayside \citep{Hoang2023} due to radiative torques (RATs) and magnetic fields (B-RATS). If radiative torques due to incident starlight or magnetic effects dominate, grains on the dayside can be oriented at an angle to the observer (in between configuration of `Side' and `Top'), an effect that is commonly used to estimate the diffuse magnetic field strength in the ISM via polarization measurement \citep[see review, ][]{Roberge2004}. This `tilting' of aerosols is also observed on Earth with turbulence causing the `wobbling' of falling ice crystals that produce elongated subsuns, and the alignment of ice crystals with electric fields above thunderclouds forming crown flashes. It is beyond the scope of this paper to explore these effects in detail, but we have included aerosol models into \texttt{POSEIDON} that can retrieve each aerosol direction separately (as shown \S \ref{POSEIDON_updates}) to test for non-mechanically aligned directional signatures in a combination that will best fit the data.

We highlight the myriad of aerosol modeling methods above to show that care must be taken when deriving aerosol absorption and scattering properties to fit exoplanet data: how lab data was taken, how directional dependent refractive indices are used to generate aerosol absorption properties, and how optical properties are derived can all change the wavelength location and morphology of an absorption feature. Additionally, temperature and polymorph transitions explored in \citet{Moran2024} can change the shape and peak wavelength of an aerosol's absorption feature. 

This work serves to highlight that potential mechanical alignment is important to consider when modeling crystalline aerosol grains, and that aerosols in exoplanet atmospheres can be used to probe atmospheric dynamics due to their coupling with winds and the geometry of the observations. Future laboratory work is slated to measure the optical properties of quartz polymorphs relevant to exoplanet atmosphere (NASA internship ID: 021096 and SEEC PI: Erika Kohler). Additionally, a JWST Cycle 4 program (JWST-GO-08309, PIs: Mullens \& Moran)
is slated to observe additional transits and eclipses of WASP-17b to constrain the quartz polymorph on the limb and potential forsterite dayside directionality. With continued efforts to measure aerosol properties in the lab and probe their spectroscopic signatures in exoplanet atmospheres, we will gain important new insights into the formation of aerosols and their interactions with the local environment. 

\subsection{\nt{\texttt{POSEIDON} V1.3.1 and Links}}\label{sec:links}

\nt{This paper coincides with the release of \texttt{POSEIDON V1.3.1} which includes a new precomputed Mie database with 144 new temperature and directional dependent aerosols included\footnote{\href{https://poseidon-retrievals.readthedocs.io/en/latest/content/opacity_database.html}{\texttt{POSEIDON}'s Opacity Database}}. It also includes the four new direcitonal slab models described in this paper (\S \ref{POSEIDON_updates}), with a tutorial notebook\footnote{For a tutorial on how to implement directionality aerosol models in \texttt{POSEIDON}, see the `Aerosol Advanced: Directionality’ tutorial in \href{https://poseidon-retrievals.readthedocs.io/en/latest/content/forward_model_tutorials.html}{Forward Model Tutorials}}.}

\nt{In this update, the effective scattering properties (asymmetry parameter, $g$, and single scattering albedo $\omega$) of all precomputed aerosols have been updated from their median values (as described in \S 2.1 of \citet{Mullens2024}) to their more accurate weighted integral values,}

\begin{equation}
g_{\mathrm{eff}} = \frac{\int \; P(r) \; \sigma_{\mathrm{geo}} \; Q_{\mathrm{scat}} \; g \; dr}{\sigma_{\mathrm{scat,eff}}}
\end{equation}

\begin{equation}
\omega_{\mathrm{eff}} = \frac{\sigma_{\mathrm{scat,eff}}}{\sigma_{\mathrm{ext,eff}}},
\end{equation}

\noindent\nt{where all variables above are defined in \S 2.1 of \citet{Mullens2024}. We have also updated the weighted single scattering albedo and asymmetry parameter utilized in the \citet{Toon1989} functions in \texttt{POSEIDON} to account for overlapping aerosols in a single pressure layer, by updating Equation 12 and 13 in \citet{Mullens2024} to}

\begin{equation}
\bar{\omega} = \frac{\Delta\tau_{Ray}}{\Delta\tau_{tot}} \; + \frac{\sum_{i}\Delta\tau_{Mie}^i \; \omega_{Mie}^i}{\Delta\tau_{tot}}
\end{equation}


\begin{align}
\bar{g} = \frac{\sum_i\omega_{Mie}^i \; \Delta\tau_{Mie}^i \; g_{Mie}^i}{\Delta\tau_{Ray} + \sum_i\omega_{Mie}^i \; \Delta\tau_{Mie}^i} 
\end{align}

\noindent\nt{where all variables above are defined in \S 2.3.1 of \citet{Mullens2024}\footnote{For a tutorial on how to implement overlapping aerosols, see the `Aerosol Advanced: Overlapping Aerosol Species + 2 Aerosol Species Patchy Clouds’ tutorial in \href{https://poseidon-retrievals.readthedocs.io/en/latest/content/forward_model_tutorials.html}{Forward Model Tutorials}}. This update was required for directional aerosol emission retrievals, as those models necessarily include overlapping aerosols in the same pressure layers. Additionally, in this update all aerosols in the precomputed databases have been precomputed with the \texttt{miepython V2.5.5}\footnote{\url{https://github.com/scottprahl/miepython}} \citep{Wiscombe1979,prahl-2024-255} package with the methodology described in \citet{Mullens2024} (where we have updated \texttt{POSEIDON} to include both \texttt{miepython} and the algorithm adapted from \texttt{LX-MIE}\footnote{\url{https://github.com/NewStrangeWorlds/LX-MIE}} \citep{KitzmannHeng2018}). While this paper utilizes the \texttt{LX-MIE} precomputed aerosol properties, the databases available for download contain the \texttt{miepython} precomputed properties. We find that both algorithms produce near-identical Mie scattering properties (up to a maximum difference of 1e-10 for $\sigma_{\mathrm{ext,eff}}$, 1e-5 for $g_{\mathrm{eff}}$, and 1e-2.5 for $\omega_{\mathrm{eff}}$), where we opt to use \texttt{miepython} since it is optimized for computational speed and can be used with multiprocessing\footnote{For a tutorial on how to precompute aerosol properties with both \texttt{miepython} and \texttt{LX-MIE}, see the `Making an Aerosol Database’ tutorial in \href{https://poseidon-retrievals.readthedocs.io/en/latest/content/forward_model_tutorials.html}{Forward Model Tutorials}}.}

\nt{The Zenodo repository for this paper (\href{https://doi.org/10.5281/zenodo.15712070}{Zenodo doi: 15712070}) contains the `Zenodo Supplementary Material' referenced in this work (which includes 1. A figure computing the cross sections of aerosols utilizing the the \citet{Fabian2001} prolate spheroid cross section derivation, 2. A figure showcasing the complete retrieval exploration of WASP-17b including the direcitonal SiO$_2$ emission retrievals, 3. A table documenting the 144 directional and temperature aerosols added to \texttt{POSEIDON}, 4. A table documenting all forward model and retrieval assumptions and priors used in this work, and 5. Nineteen figures showcasing the directional and temperature dependent aerosol cross sections versus the corresponding aerosol cross sections in the \citet{Mullens2024} aerosol database), all directional and temperature-dependent refractive indices, the precomputed aerosol database with directional and temperature-dependent aerosols, and all the code used to produce the results and figures of this work.}



\begin{acknowledgments}
We would like to express our gratitude to many people and groups for providing their insights and guidance during the development of this manuscript. We thank Dr. Esteban Gazel for his insight on silicate mineralogy. We thank Dr. Domingos De Sousa Meneses for sending his laboratory data and giving guidance on how to tie together directionality with refractive index data. We thank the curators of the Database of Optical Constants for Cosmic Dust (DOCCD) for making their directional data open-source and accessible. We thank Dr. Hannah Wakeford and Dr. Sarah Moran for preliminary comments on this work. We thank the JWST-TST-DREAMS team for their continued feedback and support. \nt{E.M. thanks Thomas Kennedy for helpful feedback on computing Mie properties.} E.M. thanks Dr. Ryan
MacDonald for help in releasing POSEIDON v1.3.1. E.M. acknowledges that this material is based upon work supported by the National Science Foundation Graduate Research Fellowship under Grant No.\ 2139899. \ntt{We thank the referee for a very constructive and thorough report of our manuscript that significantly improved the quality of this work.}
\end{acknowledgments}

%

\vspace{5mm}
\facilities{JWST(MIRI LRS)}


\software{
POSEIDON \citep{MacDonaldMadhusudhan2017, MacDonald2023},
numpy \citep[][]{harris2020array}, SciPy \citep[][]{2020SciPy-NMeth}, matplotlib \citep[][]{Hunter:2007}
}



\appendix
\S \ref{sec:crystalstructure} provides more in-depth details on crystal structures and anisotropy in crystalline aerosols. \S \ref{sec:aerosol_database} notes the new directional and temperature dependent aerosols added to the pre-computed aerosol database in \texttt{POSEIDON}. Table \ref{table:quick_check_table} is a reference table for common mineral aerosols predicted in hot Jupiters with their chemical formula, crystal system, optical properties (isotropic or anisotropic), with some notes. Figure \ref{fig:cross-sections-averages} displays the directional-dependent effective extinction cross sections ($\sigma_{\mathrm{ext,eff}}$) of SiO$_2$, Mg$_2$SiO$_4$, and MgSiO$_3$ with the two different averaging methods: the optical constant averaging method (OCA) and the spectral averaging method (SA). The two methods are both widely accepted in the literature \citep[e.g.,][]{Bohren1983,Mogli2007,Reed2017}, but result in different cross sections. Figure \ref{fig:cross-sections-temperature} displays the directional-dependent $\sigma_{\mathrm{ext,eff}}$ of SiO$_2$, Mg$_2$SiO$_4$, and MgSiO$_3$ at room temeprature, versus `hot' ($\sim$1000K), showcasing a common shift  of absorption features to longer wavelengths for hot lab samples. Figure \ref{fig:sio2_emission}, \ref{fig:mg2sio4_emission}, and \ref{fig:mgsio3_emission} are identical to Figure \ref{fig:sio2_transmission}, \ref{fig:mg2sio4_transmission}, and \ref{fig:mgsio3_transmission} but in emission (dayside) instead of transmission (limb/terminator). Figure \ref{fig:side-front} displays the relative absorption of side, front, and side + front configurations for Mg$_2$SiO$_4$ and MgSiO$_3$, assuming that if one or both configurations are dynamically favorable they would be detectable. 

\section{Crystal Structure and Anisotropy}\label{sec:crystalstructure}

Crystalline aerosols have absorption features with fine structure (`sharp' absorption features) that are more sensitive to temperature variations. Additionally, they can interact with light isotropically or anisotropically, where how light interacts with a crystalline aerosol's crystallographic axes (labeled a, b, and c in Figure \ref{fig:crystal-art}) depend on its intrinsic crystal system and resultant crystal shape through its growth habit (i.e., the atomic-level crystal lattice versus the bulk crystal shape). There are seven crystal systems (cubic, tetragonal, hexagonal, trigonal/rhombohedral, orthorhombic/rhombic, monoclinic, triclinic)
that define the length and orientation of the three crystallographic axes (an imaginary coordinate system that describe orderly arrangement inside a crystal lattice). 
From each crystal system, multiple three-dimensional crystal shapes can form depending on the growth habit. 

Cubic crystal systems have three crystallographic axes that are equal length and intersect at right angles, resulting in a bulk crystal that is optically isotropic. An example of an aerosol with a cubic crystal structure is spinel (MgAl$_2$O$_4$, octohedral crystal growth habit) \citep{Zeidler2013} and $\beta$ cristobalite (SiO$_2$, pseudo-octohedral crystal growth habit). 

Trigonal/rhombohedral crystal system have three crystallographic axes that are of equal length and are equally inclined (but not at 90 degree angles). An example of an aerosol with a trigonal crystal system is $\alpha$-quartz (SiO$_2$). When quartz crystallizes, its growth habit forms a chain of SiO$_4$ tetrahedra wind around the crystallographic c-axis (also known has the principal axis) forming a crystal shape composed of a hexagonal prism in the center and two hexagonal pyramids on the ends (see Figure \ref{fig:crystal-art}). 

Crystals with a single optical axis (an optical axis is defined as an axis in which light experiences no birefringence) are defined as uniaxial\footnote{Tetragonal, hexagonal, and trigonal/rhombohedral crystals are always uniaxial.}. Uniaxial crystals are anisotropic and have dichroic attenuation depending on two defined polarizations of light: polarizations of light parallel to the c-axis (extraordinary, A-mode, E$\parallel$c) and polarizations perpendicular to the c-axis (ordinary, E-mode, E$\perp$c)\footnote{For uniaxial crystals, the optical axis always equals the crystallographic c-axis, or the principal axis}.

Light propagating parallel to the c-axis (called `Top') will always have polarization perpendicular to the c-axis and therefore have attenuation determined by the ordinary refractive indices. Light propagating perpendicular to the c-axis (and parallel to the a-b axes, called `Side') can have polarizations perpendicular and parallel to the c-axis and therefore have attenuation determined by both ordinary and extraordinary refractive indices (see Figure \ref{fig:crystal-art}, bottom left). Because of this, when weighing the optical properties of a system of randomly oriented crystals the extraordinary ray receives a 1/3 weight while the ordinary ray receives a 2/3 weight \citep{Zeidler2013}. Uniaxial crystals can be defined as positive or negative depending on the sign of birefringence (extraordinary index - ordinary index). Quartz is a positive uniaxial crystal, meaning the c-axis is the long axis and the crystal assumes a prolate spheroid shape (in lieu of an oblate spheroid), as can be seen in Figure \ref{fig:crystal-art}. 

Forsterite (Mg$_2$SiO$_4$) and enstatite (MgSiO$_3$) both have orthrorhombic crystal systems where the three crystallographic axes are different lengths and intersect at right angles. The crystal growth habit for both forsterite and enstatite result in a tabular, euhedral shape (see Figure \ref{fig:crystal-art}, as well as \citet{Welsch2013,Beno2020}). Crystals with two optical axes are defined as biaxial crystals\footnote{Orthorhombic/rhombic, monoclinic, and triclinic are always biaxial.}. Biaxial crystals are anisotropic and have three defined sets of refractive indices that correspond to light being polarized relative to three optical directions. Optical properties are measured along each axis, and each direction gets a 1/3 weight when weighing optical properties for a system of randomly oriented crystals: E$\parallel$c for light with polarization parallel to the c-axis, E$\parallel$b for light with polarization parallel to the b-axis, and E$\parallel$a for light with polarization parallel to the a-axis\footnote{Note the relation between crystallographic axes, optical axes, and optical directions. Crystallographic axes (a,b,c) define the crystal lattice, optical axes are axes along which light experiences no birefringence (can have zero, one, or two), and optical directions are axes with different sets of refractive indices (x,y,z). For orthorhombic crystals like enstatite and forsterite, optical directions correspond to crystallographic axes (but not in a one-to-one way; for our assumptions on the relation between symmetry group, optical direction, and crystallographic axes see Table 1 in Zenodo Supplementary Material (see \S \ref{sec:links})). For monoclinic crystals, only one optical direction is parallel to the b-axis with the other two being misaligned, and for triclinic none of the optical directions correspond to the crystallographic directions. Since in this work we only deal with orthorhombic biaxial crystals, we opt to use crystallographic axis notation in lieu of optical direction notation.}.

Light propogating parallel to each axis will have attenuation determined by two sets of refractive indices; for example, light propagating along the b-axis with a polarization parallel to the c-axis will be attenuated via the E$\parallel$c refractive indices while light propagating along the b-axis with polarization parallel to the a-axis will be attenuated via the E$\parallel$a refractive indices. We refer to light propagating along the b-axis as `Front', a-axis as `Side', and c-axis as `Top' for biaxial crystals \nt{in the main text, though these designations can depend on how crystals grow relative to their optical axes, as well as how refractive indices are recorded in their original measurement references}. In Table \ref{table:quick_check_table}, we have provided common aerosols found in hot Jupiters with their crystal system, optical properties (isotropic, uniaxial, or biaxial), as well as some notes on phase transitions. 

\section{Aerosol Database Updates}\label{sec:aerosol_database}

We have updated the precomputed aerosol database featured in \citet{Mullens2024} with temperature and directional dependent cross sections of aerosols predicted to form in hot Jupiters (alongside with amorphous and polymorph variants), with a full table provided in Zenodo Supplementary Material Table 1 (see \S \ref{sec:links}). Specific species include: Hibonite \citep{Mutschke2002}, Corundum \citep{Begemann1997,Zeidler2013}, Spinel \citep{Fabian2001Spinel,Zeidler2013}, Fayalite \citep{Fabian2001}, Titanium Dioxide \citep{Posch2003,Zeidler2011}, Silicon Dioxide \citep{Zeidler2013, DeSousaMeneses2014, Moran2024}, Olivine \citep{Eckes2013,Zeidler2015}, and Orthoenstatite \citep{Zeidler2015}. Zenodo Supplementary Material Figures 3-22 (see \S \ref{sec:links}) showcases the cross sections of these aerosols compared to the aerosols in the \citet{Mullens2024} database for 0.01 \textmu m sized particles. 

Many of the refractive indices of the added aerosols only have coverage in mid-infrared (5-30 \textmu m) wavelengths, but not in UV, visible, or near-infrared wavelengths. Mid-infrared wavelengths contain composition-specific absorption cross sections of aerosol species, while shorter UVIS wavelengths are affected by (Mie) scattering. In transmission, aerosol scattering imprints itself as an increase in transit-depth towards the shortest wavelengths (i.e., a scattering slope); while in emission, aerosol scattering can generate a significant reflection signal. Recent work \citep{Grant2023,Fairman2024} has shown that jointly fitting UVIS and mid-infrared wavelengths allows for particle size, aerosol mixing ratio, and aerosol location in the atmosphere to be constrained. Since scattering properties of aerosols are more sensitive to particle size than composition, it is common practice to combine short wavelength refractive index data of a similar aerosol (same composition) with infrared data in order to generate cross sections that can be used to fit datasets with large wavelength coverage \citep[e.g.,][]{Moran2024}. The release of \texttt{POSEIDON V1.2} in \citet{Mullens2024} included a way for users to precompute aerosol properties from refractive indices and add it to the aerosol database\footnote{For a tutorial on how to precompute aerosol properties and add them to the \texttt{POSEIDON} database, see the `Making an Aerosol Database’ tutorial in \href{https://poseidon-retrievals.readthedocs.io/en/latest/content/forward_model_tutorials.html}{Forward Model Tutorials}} to allow for this sort of piecewise construction of aerosol properties.

\section{Aerosol Shape and Radiative Properties}\label{sec:aerosol-shape-appendix}
Our argument for the mechanical alignment of silicate grains in \S \ref{sec:mechanical-alignment} relies on crystalline silicates forming or growing elongated along one axis. \citet{Tsuchiyama1998} performed experiments directly condensing silicates from high temperatures and found that the resultant nanocrystals were elongated along the c-axis (conversely, amorphous particles formed at lower temperatures were found to be spherical in shape \citep{Fabian2000} \nt{however this was due to the particles experiencing a melting phase similar to the formation of near-spherical micrometeorites \citep{Kohout2014}}). \citet{Fabian2001} take into account the non-spherical shape of grains when computing the mass absorption coefficient and find that the diagnostic features of randomly oriented forsterite shift in wavelength space (notably, the 9.6 \textmu m shifts red to 9.4 \textmu m from `spheres' to `needles' while other features shift blue, see their Figure 9). There are a myriad of other methods that take into account particle shape and their particle size distributions: randomly oriented ellipsoids and continuous distribution of ellipsoids (CDE) from \citet{Fabian2000}, distribution of form factors (DFF) for aggregated spheres and Gaussian random spheres \citep{Min2006, Zeidler2013}, the rigorous numerical T-matrix method \citep{Reed2017}, axis ratios \citep{Mutschke2002}, and multiple approximations to fractal aggregate modeling (e.g., Rayleigh-Gans-Debye (RGD) theory, discrete dipole approximation (DDA), effective medium theory (EMT), distribution of hollow spheres (DHS), mean field theory (MFT), and other simplifications \citep{Tazaki2018,Lodge2024}). We have included a recreation of the prolate spheroid analysis done by \citet{Fabian2001} in our Zenodo Supplementary Material Figure 1 (see \S \ref{sec:links}) for quartz, forsterite, and enstatite and find that it can shift the absorption features, but a more in-depth exploration of non-spherical attenuation will need to be performed to to confirm these results. \nt{In our work, we used Mie theory (which assumes spherical particles) to compute aerosol opacities, which has been shown to be sufficient when comparing to lab-measured opacities, even when considering aerosol directionality \citep[e.g.,][]{Zeidler2013}.} Another factor to consider is whether thermally processed grains change their shape; meteoroids as they ablate and melt in Earth's atmosphere often form spheroid shapes and change composition due to the evaporation of metallic ions \citep{MeteorReview1,MeteorReview2}. \nt{They can, however, still retain oriented crystallographic structures and therefore interact with radiation non-isotropically, even though they are spherical in shape \citep{Kohout2014}.} We defer the thermal processing of grains to future work. 

\renewcommand{\thefigure}{A\arabic{figure}}
\setcounter{figure}{0} 
\renewcommand{\thetable}{A\arabic{table}}

\clearpage
\setlength{\extrarowheight}{1.2pt}
\startlongtable
\begin{deluxetable*}{llllll}
\colnumbers
\tablecolumns{6}
\tablewidth{0pt}
\tabletypesize{\scriptsize}
\tablecaption{Aerosol Crystal Systems and Optical Properties} \label{table:quick_check_table}
\tablehead{Aerosol& Chemical& Crystal& Optical& Notes & Example Exoplanet Reference \\
Name & Formula & System & Properties}
\startdata
Hibonite & CaAl$_{12}$O$_{19}$ & Hexagonal & Uniaxial & Condenses at high temperatures ($\sim$ 1900K). & \citet{Wakeford2017} \\
$\alpha$ Cordunum & Al$_2$O$_3$ & Trigonal & Uniaxial & Stable at high ($\sim$1400K) temperatures. & \citet{Wakeford2017}  \\
& & & & $\gamma$ is lower temperature, and $\beta$ has Na inclusions. & \citet{Moran2024}\\
Spinel & MgAl$_2$O$_4$ & Cubic & Isotropic & Condenses at high temperatures ($\sim$ 1660K). & \citet{Sudarsky2003} \\
& & & & & \citet{Wakeford2017}\\
Fayalite & Fe$_2$SiO$_4$ & Orthorhombic & Biaxial & Iron-rich olivine end member & \citet{Visscher2010}\\
Anatase & TiO$_2$ & Tetragonal & Uniaxial & Stable at room temperature & \citet{Lee2016}\\
& & & & & \citet{Moran2024}\\
Rutile & TiO$_2$ & Tetragonal & Uniaxial & Stable at high ($\sim$ 1000K) temperatures & \citet{Lee2016} \\
& & & & & \citet{Moran2024}\\
Brookite & TiO$_2$ & Orthorhombic & Biaxial & Rare polymorph at room temperature & \citet{Lee2016}\\
& & & & & \citet{Moran2024}\\
Forsterite & Mg$_2$SiO$_4$ & Orthrohombic & Biaxial & Magnesium-rich olivine end-member & \citet{Visscher2010}\\
Enstatite & MgSiO$_3$ & Orthorhombic & Biaxial & Magneisum-rich pyroxene end-member & \citet{Visscher2010} \\
& & & & Orthoenstatite is more common than clinoenstatite \\
$\alpha$ Quartz & SiO$_2$ & Trigonal & Uniaxial & Room temperature polymorph & \citet{Moran2024}\\
$\beta$ Quartz & SiO$_2$ & Trigonal & Uniaxial & Transforms from $\alpha$ quartz at $\sim$850K & \citet{Moran2024}\\
$\alpha$ Tridymite & SiO$_2$ & Orthorhombic & Biaxial & Low temperature form of $\beta$ tridymite, & \citet{Moran2024}\\
& & & & formed typically via rapid cooling/quenching.\\
$\beta$ Tridymite & SiO$_2$ & Hexagonal & Uniaxial & Transforms from $\beta$ quartz at $\sim$1150K, & \citet{Moran2024}\\
& & & & when samples have impurities.\\
$\alpha$ Cristobalite & SiO$_2$ & Tetragonal& Uniaxial & Low temperature form of $\beta$ Cristobalite,  & \citet{Moran2024}\\
& & & & formed typically via rapid cooling/quenching.\\
$\beta$ Cristobalite & SiO$_2$ & Cubic & Isotropic & Transforms from $\beta$ quartz at $\sim$1750K & \citet{Moran2024}\\
\enddata
\tablecomments{Crystalline aerosol species predicted to be present in ultra-hot and hot Jupiters with their directionalities. Aerosol/polymorph name (1), chemical formula (2), crystal system (3), optical properties (isotropic, uniaxial, or biaxial) (4), notes on aerosol species/polymorph (5), and an example exoplanet reference to that aerosol (6).}
\end{deluxetable*}

\begin{figure*}
    \includegraphics[width=0.99\textwidth]{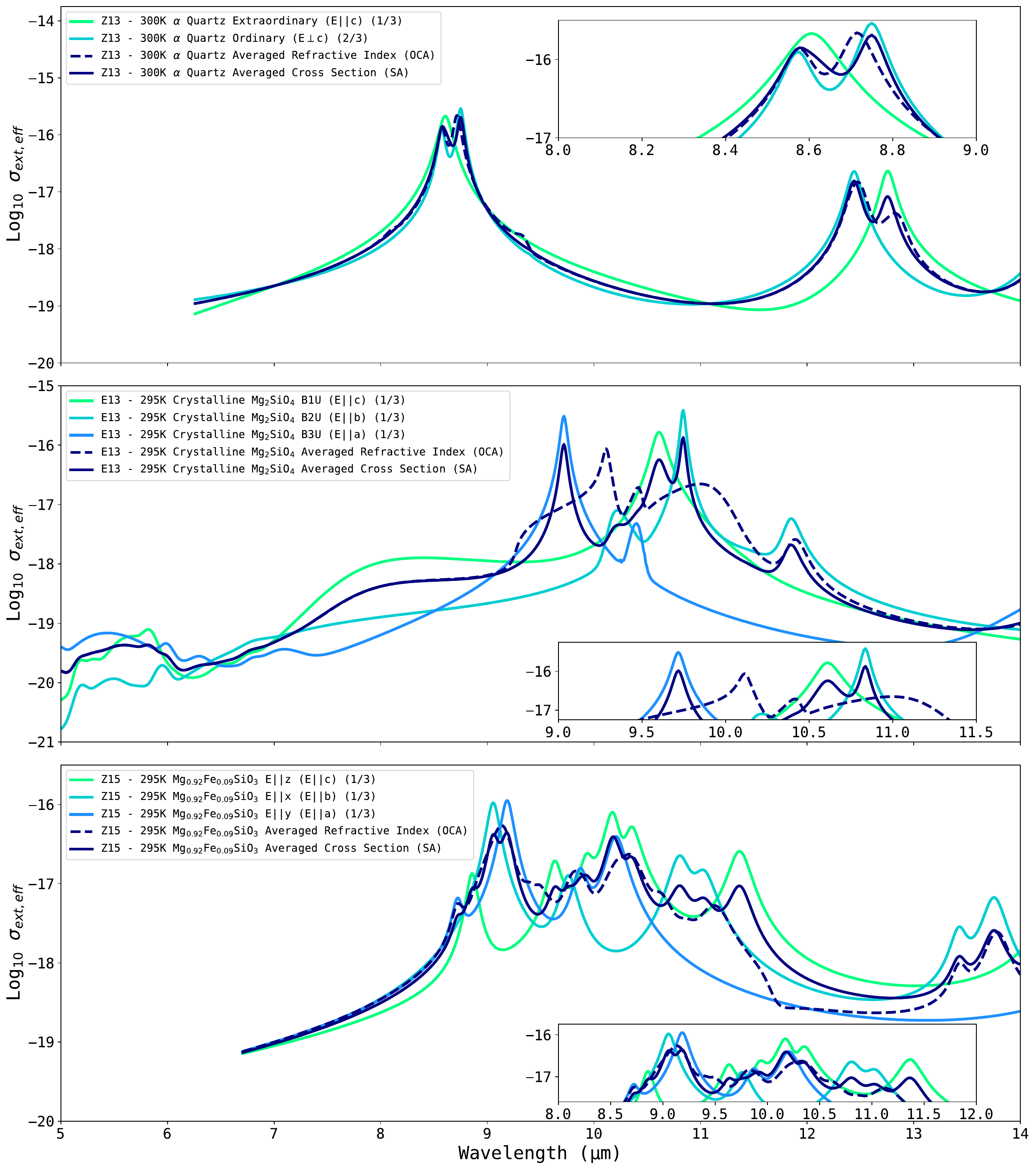}
    \caption{Same aerosol species as Figure \ref{fig:cross-sections-directionality}. Comparison of effective extinction cross sections ($\sigma_{\mathrm{ext,eff}}$, mean particle radii = 0.01 \textmu m with a lognormal distribution) computed from room-temperature, directional refractive indices vs the two methods by which to derive cross sections for randomly oriented particles. The first, averaged refractive index (optical constant averaging, OCA, dotted line), is where the directional refractive indices are averaged before computing the cross section. The second, averaged cross section (spectral averaging method, SA, solid line) is where cross sections are computed individually for each directional refractive index and then weighted. In the case of forsterite (Mg$_2$SiO$_4$), deviations between the two methods can be large. 
    }    
    \label{fig:cross-sections-averages}
\end{figure*}

\begin{figure*}
    \includegraphics[width=0.99\textwidth]{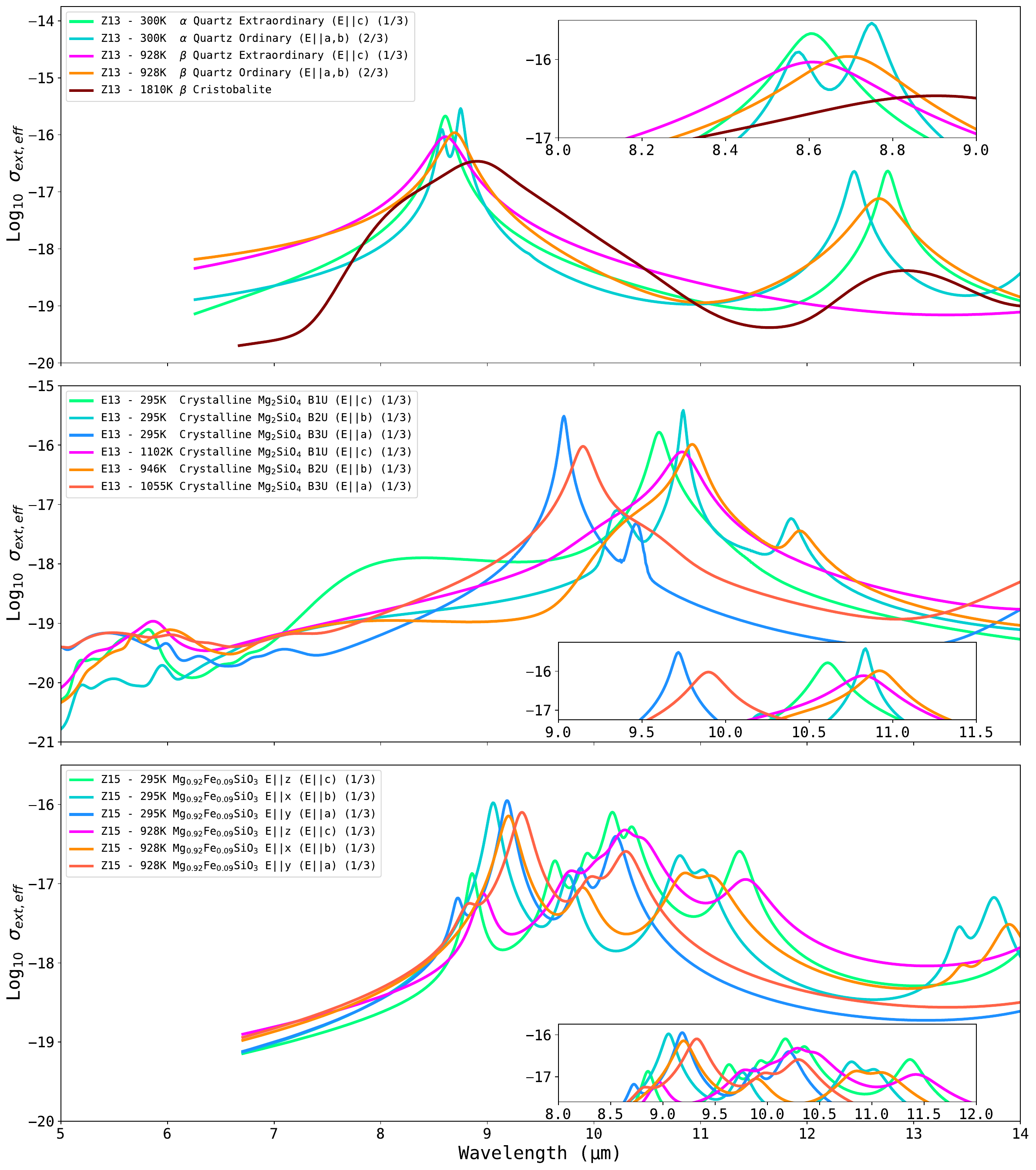}
    \caption{Comparison of effective extinction cross sections ($\sigma_{\mathrm{ext,eff}}$, mean particle radii = 0.01 \textmu m with a lognormal distribution) computed from directional, room-temperature refractive indices vs `hot' ($\sim$ 1000K) refractive indices.}    
    \label{fig:cross-sections-temperature}
\end{figure*}

\begin{figure*}
    \includegraphics[width=1.0\textwidth]{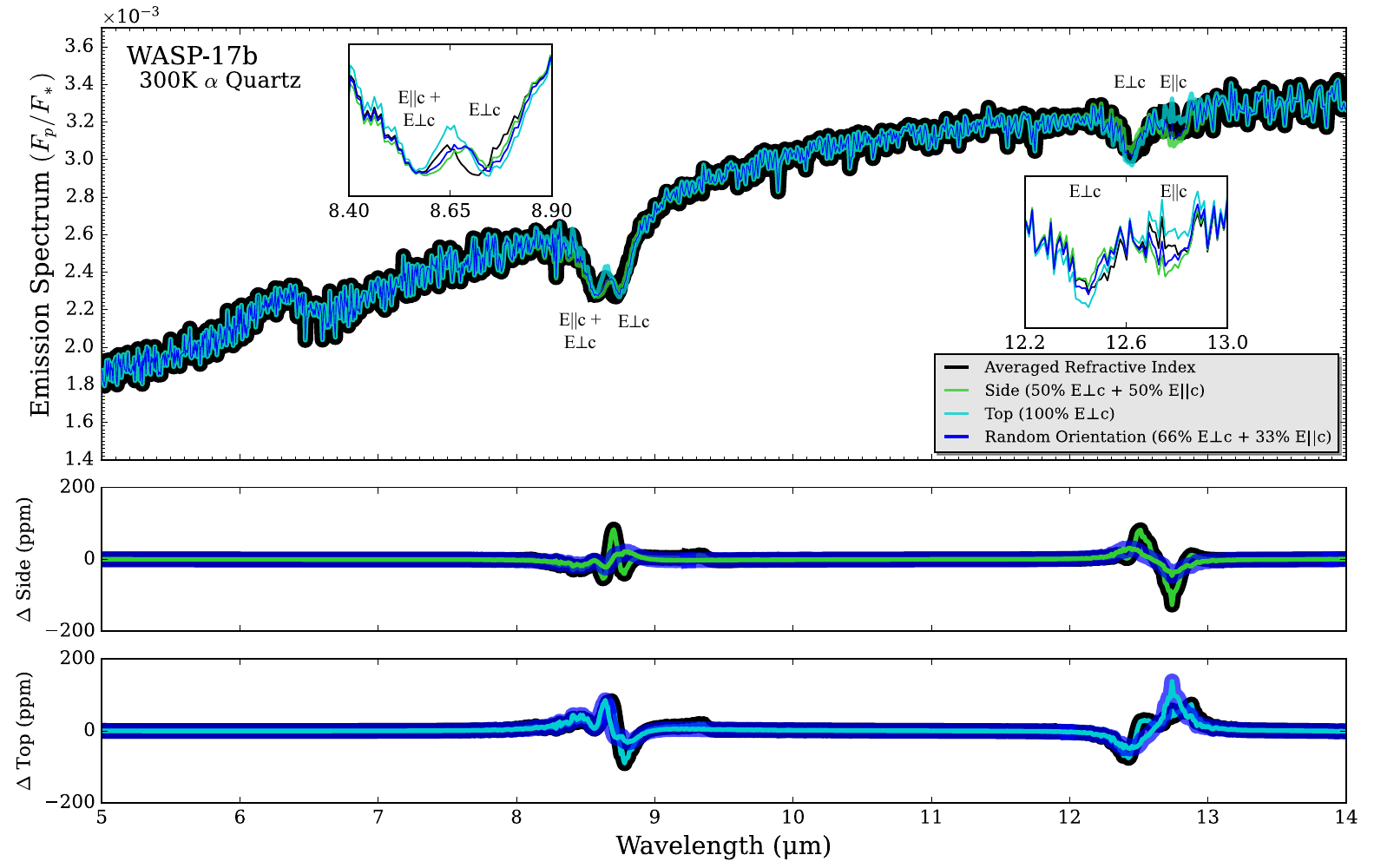}
    \caption{Same as Figure \ref{fig:sio2_transmission}, but with emission spectra.}    
    \label{fig:sio2_emission}
\end{figure*}

\begin{figure*}
    \includegraphics[width=1.0\textwidth]{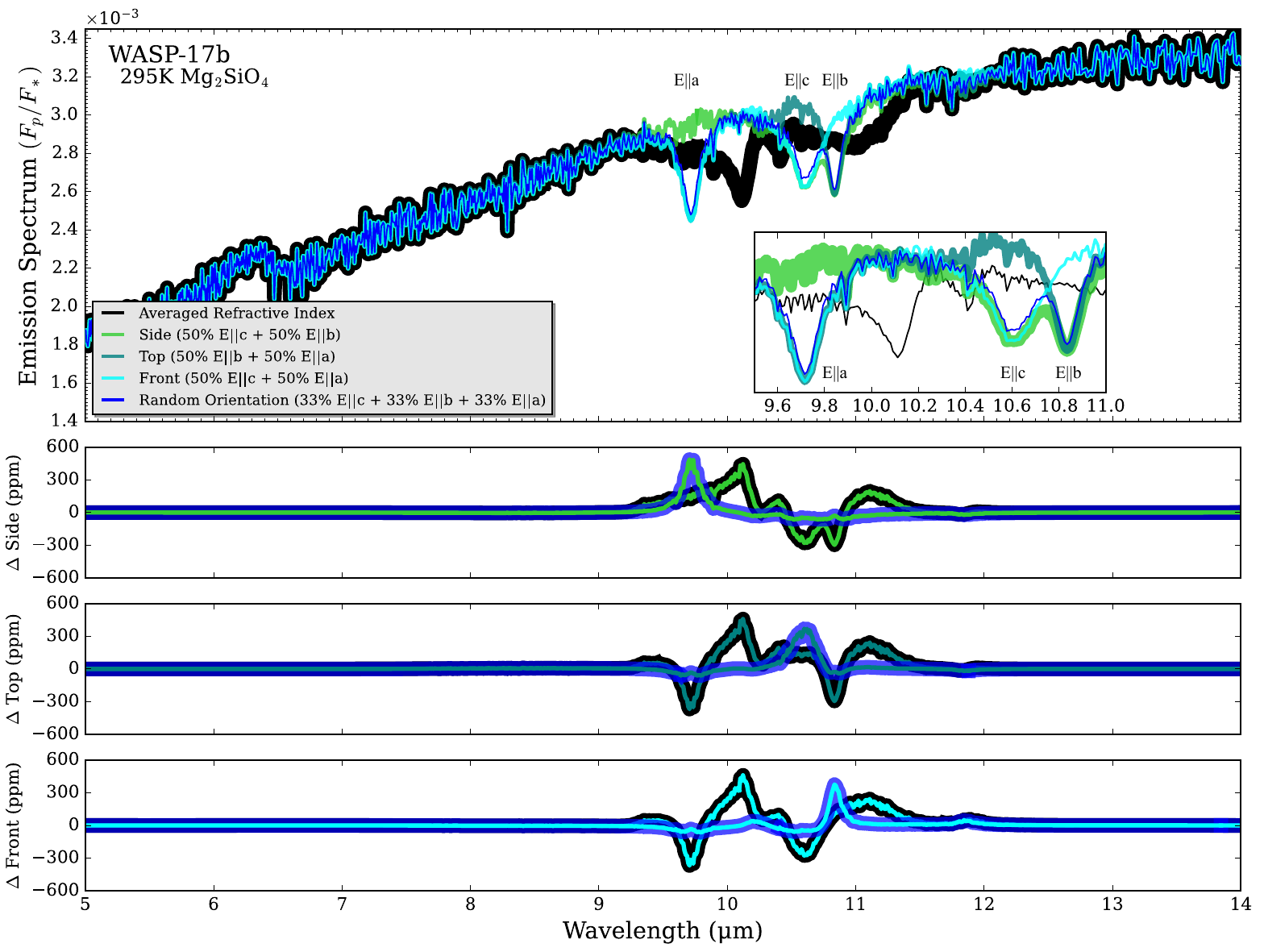}
    \caption{Same as Figure \ref{fig:mg2sio4_transmission}, but with emission spectra.}    
    \label{fig:mg2sio4_emission}
\end{figure*}

\begin{figure*}
    \includegraphics[width=1.0\textwidth]{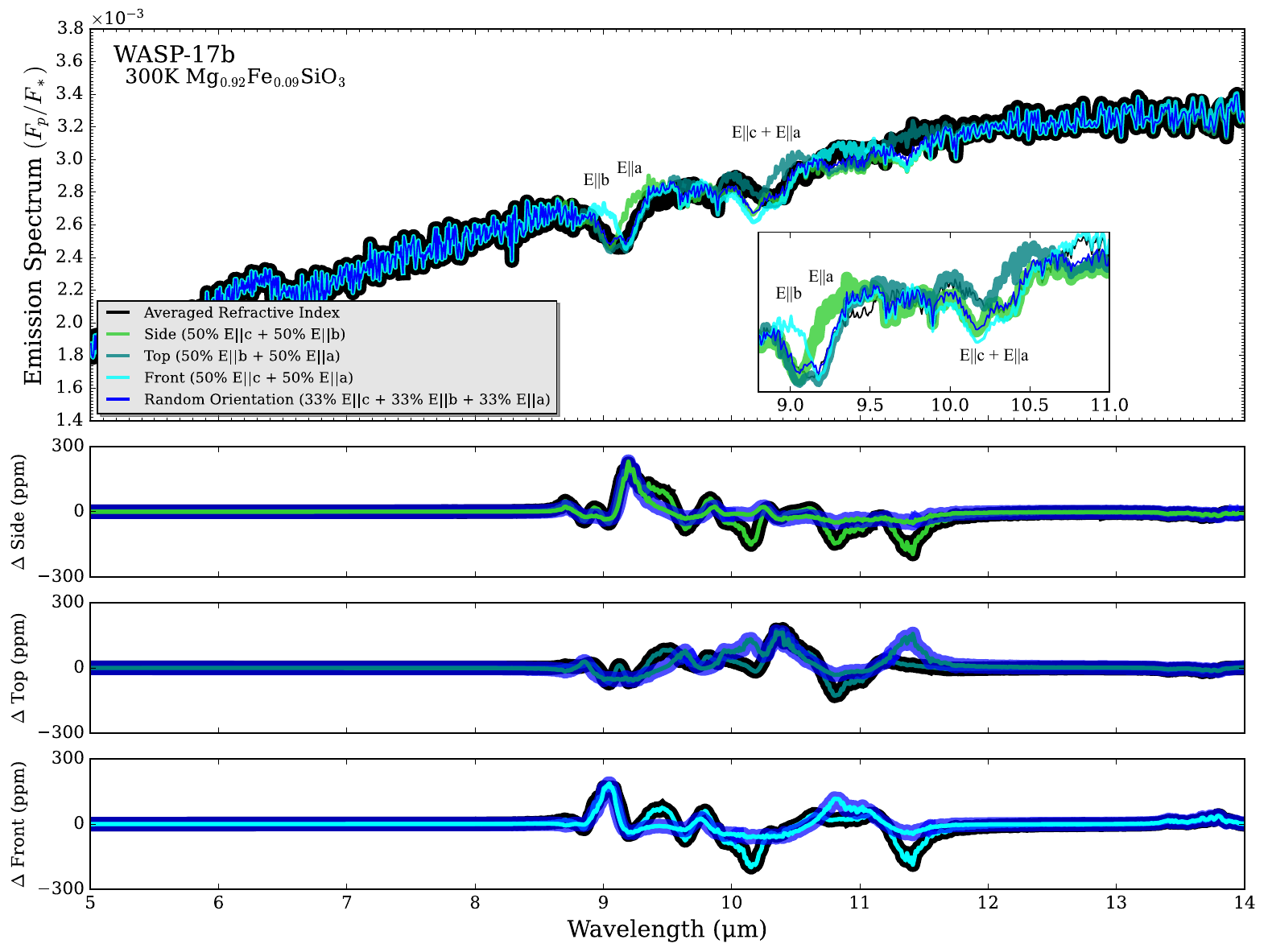}
    \caption{Same as Figure \ref{fig:mgsio3_transmission}, but with emission spectra.}       
    \label{fig:mgsio3_emission}
\end{figure*}

\begin{figure*}
    \includegraphics[width=1.0\textwidth]{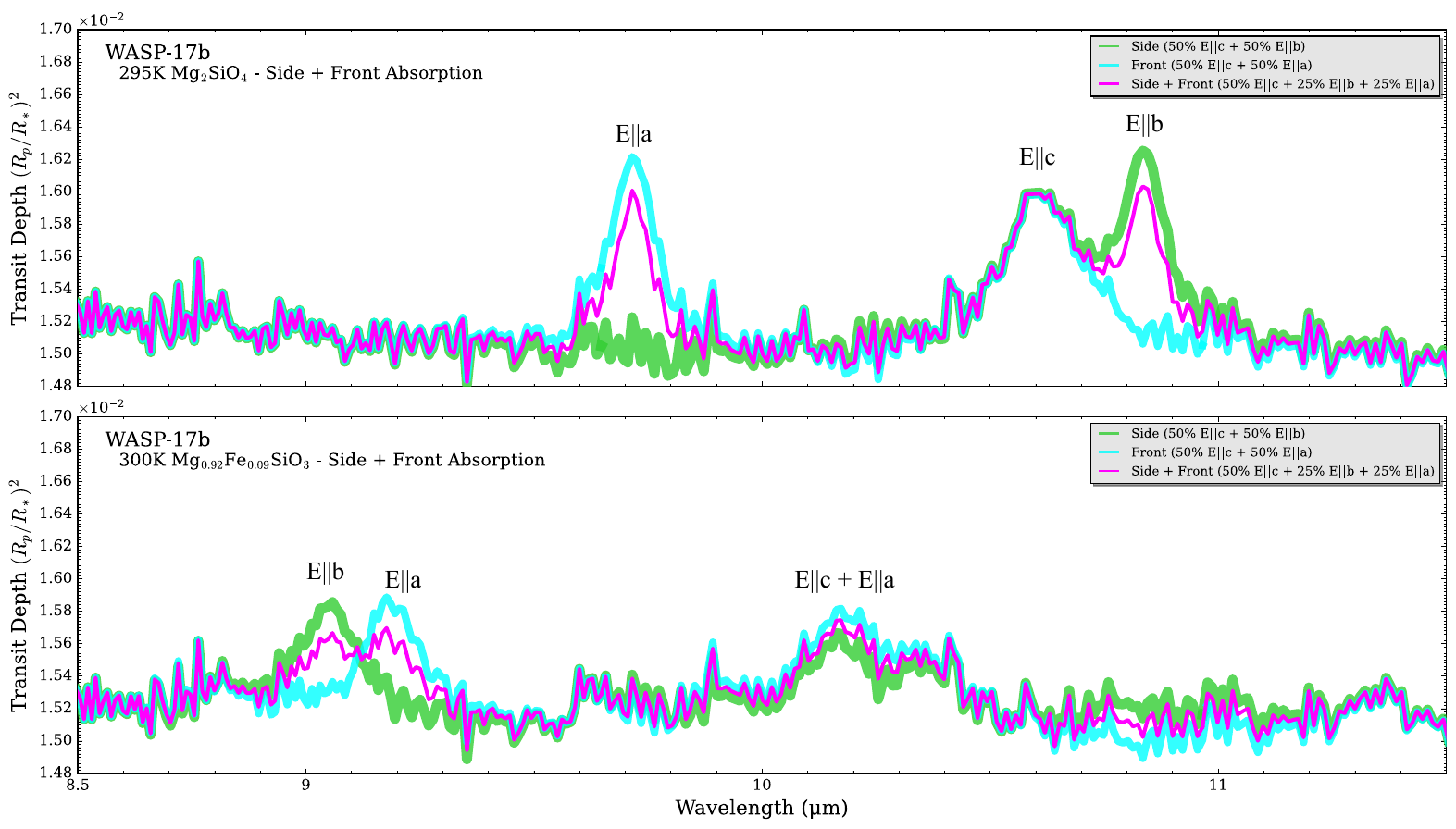}
    \caption{Showcasing relative absorption of Side, Front, and Side + Front for the biaxial crystals forsterite (Mg$_2$SiO$_4$, Top Panel) and enstatite (Mg$_{0.92}$Fe$_{0.009}$SiO$_3$, Bottom Panel). In emission, both the Side (green) and Front (cyan) configurations are permissible when assuming mechanical alignment. If one configuration is more dynamically favorable than the other, it would be detectable via a similar two out of three feature approach shown in Figures \ref{fig:mg2sio4_transmission} and \ref{fig:mgsio3_transmission}. If both are favorable, all three absorption features would be present in the data but the relative strength of the features would differ. In particular, a Side + Front configuration (magenta) would have a 50\% weighting for E$\parallel$c cross sections, and 25\% for E$\parallel$a and E$\parallel$b.}    
    \label{fig:side-front}
\end{figure*}

\clearpage
\bibliography{quartz_rainbows}{}
\bibliographystyle{aasjournal}



\end{document}